\documentclass[aps,amssymb,nofootinbib]{revtex4}
\usepackage{setspace}
\usepackage{graphicx}
\input epsf  
\epsfverbosetrue
\usepackage{psfrag}

\newcommand{\brho}{\beta^\rho}
\newcommand{\bz}{\beta^z}
\newcommand{\sigmabar}{\bar{\sigma}}
\newcommand{\omegabar}{\bar{\Omega}}
\newcommand{\rhosp}{\rho_s}
\newcommand{\rhot}{\rho_t}

\begin{document}
\title{Adaptive Mesh Refinement for Coupled Elliptic-Hyperbolic Systems}

\author{Frans Pretorius}

\address{CIAR Cosmology and Gravity Program,
         Department of Physics,
         University of Alberta, 
         Edmonton, AB T6G 2J1, \\
         Theoretical Astrophysics 130-033,
         California Institute of Technology,
         Pasadena, CA, 91125 \\}

\author{Matthew W. Choptuik}
\affiliation{CIAR Cosmology and Gravity Program \\
             Department of Physics and Astronomy,
             University of British Columbia,
             Vancouver BC, V6T 1Z1 Canada}

\begin{abstract}
We present a modification to the Berger and Oliger adaptive mesh
refinement algorithm designed to solve systems of coupled, non-linear,
hyperbolic and elliptic partial differential equations.
Such systems typically arise
during constrained evolution of the field equations of general
relativity. The novel aspect of this algorithm is a technique
of ``extrapolation and delayed solution''
used to deal with the non-local nature of the solution of 
the elliptic equations, driven by dynamical sources, within the 
usual Berger and Oliger time-stepping framework. We show
empirical results demonstrating the effectiveness of this technique 
in axisymmetric gravitational collapse simulations, and further
demonstrate that the solution time scales approximately linearly with 
problem size.
We also describe several other details of the 
code, including truncation error estimation using
a self-shadow hierarchy, and the refinement-boundary interpolation
operators that are used to help suppress spurious high-frequency
solution components (``noise'').
\end{abstract}
\maketitle

\section{Introduction}\label{sec_intro}

Adaptive mesh refinement (AMR) will be needed for many
grid-based numerical approaches designed to solve a variety of problems 
of interest in numerical relativity, including
critical gravitational collapse, binary
black hole mergers, and the study of singularity structure in
cosmological settings and black hole 
interiors \cite{lehner_review,gundlach_review,berger_review}. 
The reason is that
such problems often exhibit a wide range of relevant spatial and temporal
length scales that are impossible to adequately resolve 
with a uniform mesh, given resources available on contemporary computers.
In certain restricted scenarios, such as the head-on collision of black 
holes \cite{smarr_thesis}, or during the inspiral phase of a circular 
merger \cite{brugman_et_al,duez_et_al}, 
it is possible to construct a single, static coordinate
grid that can resolve all of the length scales. However, this requires
some {\em a priori} knowledge of the structure of the solution that will 
not be available in general.
To date, in numerical relativity, mesh refinement has been used quite effectively in 1D and 2D 
critical collapse simulations \cite{choptuik,hamada_stewart,hamada_et_al,
choptuik_et_al,liebling_choptuik,chl,paper2,paper3}, the study of critical phenomena
in the nonlinear sigma model in 3D Minkowski space \cite{liebling}, 
in 2D simulations of cosmological
spacetimes \cite{hern}, 3D simulation of gravitational 
waves and single black holes\cite{brugmann,papadopoulos_et_al,new_et_al,
schnetter_et_al,imbiriba_et_al,gh3d_p1}, 
to construct initial data for binary black hole mergers
\cite{diener_et_al,brown_lowe,brown_lowe2}, and the evolution of binary black hole
spacetimes \cite{brugman_et_al,gh3d_p2}\footnote{The
notation 1D refers to problems with dependence on 1 spatial dimension, in addition to the implicit
dependence on
time; similarly for 2D and 3D.}. 

In the following we only consider Cauchy evolution; in other words, we have a 
timelike coordinate $t$ that foliates the spacetime into a set of spacelike 
slices, and, given initial data at $t=0$, we want to evolve to the future $t>0$.
In such a coordinate basis the field equations of general relativity 
consist of a set of 10, second order, quasi-linear
partial differential equations (PDEs) for 10 metric coefficients that describe
the structure of spacetime. Thus, from a Lagrangian perspective,
one would expect ten dynamical degrees of freedom per point in spacetime,
where each degree of freedom is specified by a pair of values---a generalized 
coordinate and its conjugate momenta (here, for example, a metric element
and its first time derivative). However, four of the field equations
do not contain second time derivatives of the metric, and therefore serve
as {\em constraints}, eliminating four dynamical degrees of freedom
(these four equations are usually called the {\em constraint equations},
and the six remaining equations the {\em evolution equations}).
Furthermore, the geometry of spacetime is
invariant under arbitrary coordinate transformations of the 4 spacetime 
coordinates, and choosing a particular coordinate system (or ``gauge'') amounts to
imposing four additional constraints, leaving 
only two dynamical degrees of freedom per spacetime point.

The presence of constraints and coordinate freedom in the Einstein equations permits
considerable leeway in the solution method. One of the more common methods used
these days is so-called {\em free evolution} within the ADM (Arnowitt-Deser-Misner\cite{ADM})
decomposition (see ~\cite{TP-80} for a thorough discussion of the
various possibilities and corresponding classification scheme).
Here, the 4 dimensional spacetime metric is written as a 3 dimensional
spatial metric (6 independent components), {\em lapse function} and a 
spatial {\em shift vector} (3 components)~\footnote{
We emphasize that our counting is of second-order-in-space-and-time equations.  
A common approach is to recast the 6 second order equations into a system
of first order equations, which result in 
additional auxiliary quantities that must be evolved, and additional
constraints among the new quantities.}.
A coordinate system is chosen by specifying the lapse and shift, the
constraint equations are solved at the initial time, and
the spatial metric is then evolved in time using the evolution equations.
In the continuum limit, 
the Bianchi identities (see \cite{MTW}, for example) guarantee
that such an evolution scheme will preserve the constraints for all
time, given appropriate boundary conditions. 

The constraint equations are elliptic in nature, whereas the evolution 
equations are hyperbolic. Coordinate conditions can be chosen that
give algebraic, elliptic, parabolic or hyperbolic equations for 
the kinematical variables (lapse and shift). Thus, even though
it is always possible in principle to adopt a free evolution approach
in the numerical solution of Einstein's equations where elliptic equations
are only solved at the initial time, there are nonetheless two situations
where it may be preferable or necessary to solve one or more elliptic equations at
each time step of the evolution. 
First, as just mentioned, it may be useful to adopt elliptic 
coordinate conditions.
For example the choice of {\em maximal slicing} yields an elliptic equation for the 
lapse function, and the {\em minimal distortion} condition gives a set
of elliptic equations for the shift vector components~\cite{york_78}.
 
Second, in a numerical evolution, since the 
Bianchi identities will only be satisfied to within truncation
error, the constraints can only be preserved to within truncation error~\cite{MWC-91}.
This is not necessarily problematic, as the violation of the constraints
should converge away in any consistent, stable numerical code.
However, studies have indicated that with some formulations of the field
equations, and using free evolution, certain constraint-violating 
modes grow exponentially with time, requiring prohibitively high accuracy
(hence resolution) of the initial data and subsequent evolution
to obtain a solution that is sufficiently close to the
continuum one for the desired length of time 
integration\cite{lindblom_scheel,shinkai_et_al,lehner_review}.
One possible method to circumvent this problem is to use 
{\em constrained evolution} rather than free evolution; in this 
case one uses some number, $m$,  $m\le4$, of the constraint equations 
to fix $m$ of the dynamical variables, in lieu of the $m$ second-order 
in time equations that would otherwise be used to update those 
quantities~\footnote{An alternative approach that has recently begun to
receive some attention is {\em constraint-projection} (also called
{\em chopped evolution} in \cite{TP-80}), where
periodically during a free evolution the constraints are re-solved
using a subset of the evolved solution supplied as free data for the
constraint solve, and afterward the evolution is continued 
with this new ``initial data'' \cite{anderson_matzner,matzner,holst_et_al}}. 
Again, the expectation is that a stable numerical
code will, in the continuum limit, provide a solution that is consistent
with {\em all} of the field equations, including, in this case, those 
evolution equations that are not explicitly used in the overall update
scheme.  (Additionally, one hopes that violations in the evolution equations
will not grow exponentially in time).
To date, a 
significant majority of numerical codes use free evolution\cite{lehner_review}, 
and with the exception of \cite{bonazzola_et_al} all constrained evolution
simulations have been carried out in 1D or 2D. \footnote{We should also mention here that there
are alternative methods for evolving the field equation other than
those based on a 3+1 decomposition (see \cite{lehner_review}); in particular,
characteristic or null evolution has proven useful in many situations,
and recently an AMR scheme for characteristic evolution has been proposed
\cite{null_amr}.
}

At this point, it is worth noting that there apparently is still
an impression in the relativity community as a whole that elliptic equations 
are computationally expensive, and are thus to be avoided in numerical evolution if 
at all possible.  However, provided that appropriate algorithms---such as the 
multigrid method---are adopted, this view is not consistent with at least some 
experience (see also \cite{pfeiffer_et_al} for fast elliptic solvers using spectral
methods). In particular, as we will show below, the solution of the elliptic equations
in our adaptive code requires roughly twice the CPU time
required to solve the hyperbolic equations; furthermore, the solution
cost of the elliptics scale {\em linearly} with the size of the computational domain.
Thus, if one considers that in a free evolution an equivalent number of hyperbolic
equations would need to be solved in lieu of the elliptic equations, the difference
in cost is {\em not} a significant issue in deciding whether to tackle a 
particular problem using constrained versus free evolution. 

We are thus lead to consider AMR algorithms for mixed elliptic/hyperbolic 
type, where our use of the term ``hyperbolic'' does not denote any formal
definition of hyperbolicity, but instead is used to refer to an equation 
that is characterized by locality of influence. 
Now, a very well known AMR algorithm for hyperbolic equations
is due to Berger and Oliger (B\&O)~\cite{BO}. This algorithm
has several important properties that make it quite useful and efficient in 
solving certain classes of problems. These properties include dynamical
regridding via local truncation error (TE) estimates, a grid-hierarchy
composed of unigrid (single mesh) building blocks, and a recursive
time-stepping algorithm that provides ``optimal'' efficiency in solving
discretized evolution equations that are subject to a CFL-type stability condition.
However, this algorithm, implemented verbatim for a mixed elliptic/hyperbolic
system {\em cannot} be expected to work in general, due in 
part to the non-local nature of elliptic equations, and
in part to the non-linear nature of the elliptic equations that tend 
to arise in numerical relativity. The reasons 
for this are as follows (a more detailed discussion is given in Sec.~\ref{sec_orig_BO}
and Sec.~\ref{sec_mod_BO}). In the B\&O time-stepping procedure, a single,
large time step is taken on a coarser level before several smaller 
time steps are taken on the interior, fine level. This is done so 
that the solution obtained on the coarse level can be used to set
boundary conditions (via interpolation in time) for the subsequent
fine level evolution. As the hierarchy is generated via local
truncation error estimates, the solution obtained on the coarse
level in the vicinity of the fine level boundaries will presumably
be sufficiently accurate to allow one to use the coarse
level solution to set fine level boundary conditions without adversely affecting
the global solution. If the equations are hyperbolic, then a 
poorly resolved solution in the interior region of the coarse
level will not have time to ``pollute'' the coarse/fine boundary region
in only a single coarse level evolution step (and the coarse level solution
is refreshed every time step with the fine level solution in the injection
phase of the algorithm).
This last statement is not true for elliptic equations in general, for then poorly
resolved source functions in the interior of the coarse level
could globally affect the accuracy of the solution obtained during
the coarse level evolution step. For certain kinds of linear elliptic equations,
such as the Poisson equation in Newtonian gravity, or that arising
in the incompressible Navier-Stokes equations, one can circumvent
this problem by taking advantage of conservation laws satisfied by
source fields that couple to the elliptic equtions (for example, 
with the Poisson equation in Newtonian gravity one can use a fine to 
coarse level injection function that preserves the matter energy density,
and see for example \cite{almgren_et_al} for a method for solving
the Navier-Stokes equations with B\&O style time sub-cycling).
In general relativity the equations are non-linear, and 
furthermore are coupled in such a manner that it is impossible to isolate 
such source functions in general.
Therefore, to use B\&O AMR in a constrained evolution,
in particular its time-stepping algorithm, requires some modifications
to deal with the elliptic equations; these modifications are the prime 
focus of this paper.

In Sec.~\ref{sec_graxi}, we first review the axisymmetric gravitational collapse
code introduced in \cite{paper1} that was used to develop the AMR 
approach described in this paper. This code solves a discretized version
of the Einstein-Klein Gordon system of equations. We then review the original 
Berger and Oliger algorithm in Sec.~\ref{sec_orig_BO}, and then 
proceed to a description of the modifications we have made to handle 
elliptic equations in Sec.~\ref{sec_mod_BO}.
Our primary modification
involves a split of the solution of the elliptic equations into two
phases. During the first phase, when hyperbolic equations
are solved, functions satisfying elliptic equations are extrapolated
to the advanced time level. The second phase is delayed until all finer levels 
have been evolved in the same fashion via recursion, and hence all levels
at the same or finer resolution as the given one are in sync (i.e.~have been 
evolved to the same physical time). Then,
the elliptic equations are solved over the {\em entire} sub-hierarchy
from the given level to the finest, using {\em extrapolated} boundary conditions
from the parent (coarser) level at interior coarse-grid boundaries.
In Sec.~\ref{sec_results} we present several
simulation results, including convergence tests and comparison 
with unigrid simulations. Concluding remarks are given in 
Sec.~\ref{sec_conclusion}. All equations and finite-difference operators
are listed in App. \ref{appendix}, and some additional 
details of the AMR algorithm are given in App. \ref{algorithm_details}.

\section{An Axisymmetric Gravitational Collapse Code}\label{sec_graxi}

In this section we briefly review the physical system we are 
modeling (general relativity with a scalar field matter source), the PDEs governing the
model, and the 
unigrid numerical code that computes an approximate finite-difference solution of 
the PDEs;
additional details can be 
found in \cite{paper1,fpthesis}.

\subsection{Equations, Coordinate System and Variables}

The Einstein field equations can be written as
\begin{equation}\label{einstein_tens}
R_{\mu\nu} - \frac{1}{2} R g_{\mu\nu} = 8\pi T_{\mu\nu},
\end{equation}
where $R_{\mu\nu}$ is the Ricci tensor, $R\equiv R^\mu{}_\mu$ is
the Ricci scalar (using the Einstein summation
convention where repeated indices are summed over), 
and we use geometric units where
Newton's constant, $G$, and the speed of light, $c$, are set to 1
\cite{MTW}.
With a massless scalar field $\Phi$ as the matter source,
the stress-energy tensor $T_{\mu\nu}$ is given by
\begin{equation}\label{set}
T_{\mu\nu} = 2 \Phi_{,\mu}\Phi_{,\nu} - g_{\mu\nu} \Phi_{,\gamma}\Phi^{,\gamma},
\end{equation}
and the evolution of $\Phi$ is governed by the wave equation
\begin{equation}\label{phi_eom}
\Box \Phi \equiv \Phi_{;\mu}{}^\mu = 0.
\end{equation}
In these expressions a comma (,) is used to denote a partial derivative,
and a semicolon (;) a covariant derivative.

Restricting attention to axisymmetric spacetimes without angular momentum,
and choosing cylindrical coordinates, $(t,\rho,z,\phi)$, adapted to the
symmetry, we can write the spacetime metric as
\begin{equation}
ds^2 = -\alpha^2 dt^2
       + \psi^4 \bigl[   \,   (d\rho + \brho dt)^2
                       + (dz + \bz dt)^2
                       + \rho^2 e^{2 \rho\sigmabar} d\phi^2
                \bigr].
\label{metric}
\end{equation}
The axial Killing vector is $(\partial/\partial \phi)^\mu$ and hence
all the metric functions $\alpha,\brho,\bz,\psi$ and $\sigmabar$, and
the scalar field $\Phi$ depend only on $\rho,z$ and $t$. Almost all
coordinate freedom has been eliminated by choosing this form for the
metric. What remains to be specified is a time-slicing, and for this
we use maximal slicing, defined by
\begin{equation}
K=0, \ \ \ \ \ \ \frac{\partial K}{\partial t}=0,
\end{equation}
where $K\equiv K_a{}^a$ is the trace of the extrinsic curvature
tensor $K_a{}^b$ of $t={\rm const.}$ slices \cite{york_78}. This gives
an elliptic equation for $\alpha$ (\ref{slicing_eqn}).
The constraint equation subset of (\ref{einstein_tens}) 
gives 3 additional elliptic equations: the {\em Hamiltonian constraint},
which is viewed as an equation for $\psi$ (\ref{hc_eqn}), and the $\rho$ and $z$ components 
of the {\em momentum constraint}, which are treated as equations  for $\brho$ (\ref{brho_eqn}) and 
$\bz$ (\ref{bz_eqn}), respectively. One member of the evolution 
subset of (\ref{einstein_tens})
yields a second order evolution equation for $\sigmabar$, 
and the wave equation (\ref{phi_eom}) provides a second order hyperbolic equation for $\Phi$. We convert 
both of these evolution equations to first-order-in time-form (\ref{omegabar_eqn},\ref{pi_phi_eqn})
by defining conjugate variables $\omegabar$ (geometry) and $\Pi$ (matter) as follows:

\begin{equation}\label{omegabar_def}
\alpha\bar{\Omega} = - \sigmabar_{,t}
     +  2 \beta^{\rho} \left( \rho\sigmabar \right)_{,\rho^2} + \beta^z
        \sigmabar_{,z}
     - \left[ {\beta^{\rho} \over \rho} \right]_{,\rho}
\end{equation}

\begin{equation}\label{Pi_def}
\Pi \equiv \frac{\psi^2}{\alpha} \left(\Phi_{,t} - \brho \Phi_{,\rho}
                                       + \bz \Phi_{,z}\right).
\end{equation}
Thus we end up with a system of $8$ equations for $8$ 
variables---$\alpha,\psi,\brho$ and $\bz$ satisfy elliptic
equations, and $\sigmabar,\omegabar,\Phi$ and $\Pi$ satisfy 
hyperbolic equations.

\subsection{Boundary Conditions}
On the axis at $\rho=0$, the following regularity conditions must be enforced in order 
that spacetime remain locally flat in the vicinity of the axis:
\begin{eqnarray}
\alpha_{,\rho} & = & 0, \nonumber\\
\psi_{,\rho} & = & 0, \nonumber\\
\bz_{,\rho} & = & 0, \nonumber\\
\brho & = & 0, \nonumber\\
\sigmabar & = & 0, \nonumber\\
\omegabar & = & 0, \nonumber\\
\Phi_{,\rho} & = & 0, \ \ \ \nonumber\\
\Pi_{,\rho} & = & 0. \label{ibc}
\end{eqnarray}
At the outer boundaries $\rho=\rho_{max}$, $z=z_{max}$ and $z=z_{min}$,
for the hyperbolic variables approximate outgoing radiation (Sommerfeld)
conditions are imposed~\footnote{
These conditions assume that spacetime is nearly flat at the outer boundary, 
and that disturbances (radiation) in both the scalar and gravitational fields
are propagating purely radially, and have $1/r$ falloff.
}, while for the elliptic equations,  conditions based on asymptotic
flatness conditions are used:
\begin{eqnarray}
\alpha &=& 1 , \nonumber\\
\psi-1 + \rho \psi_{,\rho} + z \psi_{,z} &=& 0 , \nonumber\\
\bz &=& 0 ,
\nonumber\\
\brho &=& 0 , \nonumber\\
r \sigmabar_{,t} + \rho \sigmabar_{,\rho} + z \sigmabar_{,z} + \sigmabar&=&0
, \nonumber\\
r \omegabar_{,t} + \rho \omegabar_{,\rho} + z \omegabar_{,z} + \omegabar&=&0
, \nonumber\\
r \Phi_{,t} + \rho \Phi_{,\rho} + z \Phi_{,z} + \Phi&=&0
, \nonumber\\
r \Pi_{,t} + \rho \Pi_{,\rho} + z \Pi_{,z} + \Pi&=&0.
\label{obc}
\end{eqnarray}

\subsection{Unigrid Numerical Scheme}\label{sec_unigrid}

The set of $8$ PDEs are solved using
second-order accurate finite difference (FD) techniques.
The elliptic equations are solved using an FAS (Full 
Approximation Storage)
multigrid algorithm with $V$-cycling \cite{brandt,MG}. 
At $t=0$, $\sigmabar,\omegabar,\Phi$
and $\Pi$ are freely specified, after which the remaining variables
$\alpha,\brho,\bz$ and $\psi$ are obtained by solving the corresponding
elliptic equations.
To evolve the variables with time, the hyperbolic equations are 
discretized using an iterative Crank-Nicholson scheme, 
with Kreiss-Oliger\cite{KO} dissipation added to reduce unwanted 
(and un-physical) high-frequency solution components (``noise'') 
of the FD equations. This iteration involves variables
at two time levels: the known solution at $t=t_0$, and the
unknowns, solved for using Newton-Gauss-Seidel relaxation
implemented in RNPL \cite{rnpl}, at $t=t_1=t_0+\Delta t$. After
each iteration, the elliptic variables are updated at $t=t_1$ by
applying a single $V$-cycle. This process is repeated until the
infinity norm of the residual of all equations is below some specified
tolerance. The pseudo-code in Fig.~\ref{ngs_iteration} summarizes this iteration sequence:

\begin{figure}
\epsfxsize=17cm
\begin{flushleft}
\begin{obeylines}
{\tt
~
~~~~~~As an initial guess to the solution at time $t_1$, copy variables
~~~~~~~~~from $t_0$ to $t_1$;
~
~~~~~~repeat
~~~~~~~~~perform 1 Newton-Gauss-Seidel relaxation sweep of the
~~~~~~~~~~~~evolution equations, solving for the unknowns at time $t_1$;
~~~~~~~~~perform 1 multigrid vcycle on the set of elliptic equations,
~~~~~~~~~~~~discretized at time $t_1$;
~~~~~~until (residual norm < tolerance)
~
}
\end{obeylines}
\end{flushleft}
\caption
{A pseudo-code description of the iteration we use to solve the system of
coupled hyperbolic/elliptic equations on a single mesh.
\label{ngs_iteration} }
\end{figure}

Specific difference operators used to discretize the equations
are summarized in App. \ref{app_fde}.

\section{The Berger and Oliger AMR Algorithm}\label{sec_orig_BO}

Here we briefly review some aspects of the B\&O AMR algorithm for
hyperbolic PDEs that are of relevance to this paper, in particular 
the grid hierarchy and time-stepping procedure.

\subsection{AMR Grid Hierarchy}

\begin{figure}
\begin{center}
\includegraphics[width=16cm,clip=true,viewport=50 250 595 842]{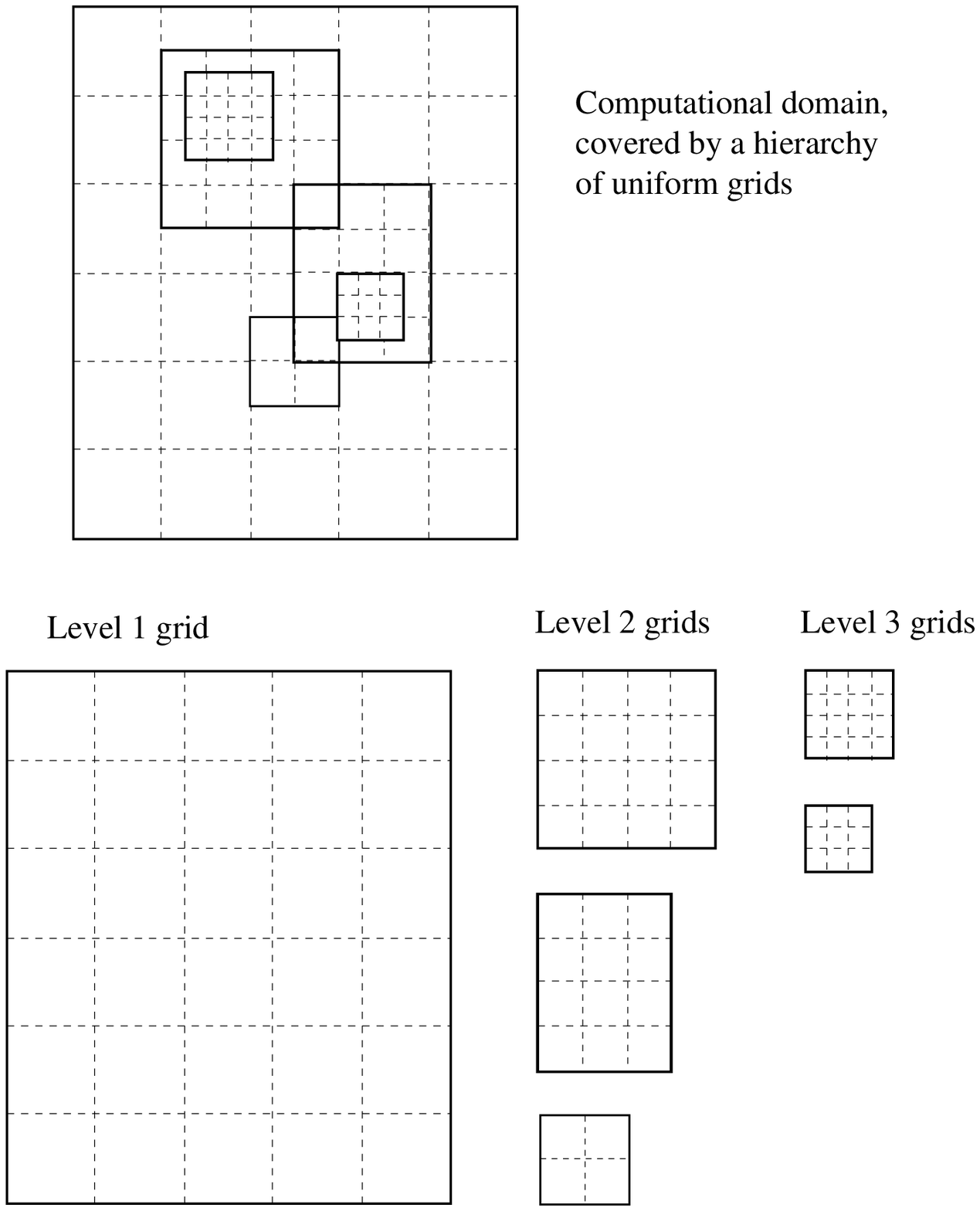}
\end{center}
\caption
{An example of a Berger and Oliger mesh hierarchy. The hierarchy consists
of a number of {\em levels}, where each level contains a set of uniform meshes
of the same spatial resolution. Our convention is to let higher level numbers
denote levels consisting of grids with higher resolution (smaller
mesh spacing, $h$), beginning with level 1 for the coarsest mesh.
The upper diagram shows the computational domain, covered by a
three-level-deep hierarchy. The plots below this
demonstrate how the hierarchy is stored in memory, namely as a collection
of individual grids. Thus a given point, ${\vec x}$ in the computational domain
can be contained/represented in multiple grids in the Berger and Oliger
scheme.
\label{BO_mesh_struct}}
\end{figure}

In the Berger and Oliger AMR algorithm, the computational domain is
decomposed into a hierarchy of uniform meshes (see Fig.~\ref{BO_mesh_struct})
with the following properties:
\begin{itemize}
\item The hierarchy contains $\ell_f$ {\em levels}. Each level $\ell$ contains
grids of the same resolution---the coarsest grids are in level $1$ (the
{\em base grid}), the next-coarsest in level $2$, and so on until
level $\ell_f$, which contains the finest grids in the hierarchy.
\item The ratio of discretization scales $h_{\ell}/h_{\ell+1}$ between
levels $\ell$ and $\ell+1$ is called the {\em spatial refinement ratio}
$\rho_{s,\ell}$. $\rho_{s,\ell}$ is typically an integer greater than
or equal to 2. For simplicity, we will also assume that $\rho_{s,\ell}$
is the same for all levels,
and therefore use the symbol $\rhosp$ to denote the
spatial refinement ratio.
\item All grids at level $\ell+1$ ({\em child} grids) are {\em entirely}
contained within grids at level $\ell$ ({\em parent} grids). Grids
at the same level may overlap.
\item In the simplified variant of the B\&O algorithm described here, 
we require that all grids within the hierarchy share the same coordinate system.
In particular, this implies that all grid boundaries run parallel to the
corresponding boundaries of the computational domain.
In addition, a child grid must be aligned relative to its
parent grid such that all points on the parent grid, within the common overlap
region, are coincident with a point on the child level. The original
B\&O algorithm allowed for a child grid to be rotated relative to its
parent.
\end{itemize}

The particular grid structure that exists at any given time is calculated 
by computing local truncation error estimates,
so that at any point $\vec{x}\equiv(\rho,z)$ within the computational domain
the finest grid covering that point has sufficient resolution
to adequately resolve all features of the solution there.
This is an important property of the grid hierarchy, not only for the
obvious reason of providing the desired resolution everywhere, but
it justifies the use of the B\&O time-stepping algorithm to evolve
the hierarchy, as we now review.

\subsection{The Berger and Oliger Time-Stepping Algorithm}

The B\&O recursive time-stepping algorithm was designed to
solve hyperbolic equations, discretized on the AMR grid hierarchy.
The basic ideas behind this update scheme are as follows.
The hierarchy is evolved in time through a particular sequence
of unigrid time-steps, performed on individual grids within the
hierarchy. A time step of size $\Delta t_\ell$ is taken on all
grids at level $\ell$, {\em before} a number $\rho_{t,\ell}$ (the {\em temporal
refinement ratio}) time steps of size $\Delta t_{\ell+1}=
\Delta t_\ell/\rho_{t,\ell}$
are taken on level $\ell+1$. The preceding rule is applied recursively,
from the coarsest to finest level in the hierarchy. In general, $\rho_{t,\ell}$
must be an integer greater than or equal to  $\rho_{s,\ell}$
in order to satisfy the CFL condition on
all levels in the hierarchy if it is satisfied on the coarsest level.
As with $\rhosp$, we only consider a constant temporal refinement
ratio $\rhot$ for all levels.
The reason why a time step is first taken on coarse level $\ell$
is that the solution obtained there at time $t+\Delta t_\ell$ is then
used to set boundary conditions,
via interpolation in time, for the subsequent time steps on the finer
level $\ell+1$ (unless some portion of the fine level abuts the boundary
of the computational domain, in which case the physical/mathematical 
boundary conditions of the original problem can be applied.)
It is possible to do this because the solution
obtained on the coarse level in the vicinity of the
finer level boundary will be {\em as accurate}, to within the
specified truncation error, as a putative solution would have been that was
obtained on a fine level encompassing the entire computational 
domain\footnote{Assuming that the specified maximum TE estimate 
used in the construction of the grid hierarchy is sufficiently small that the solution
is within the convergent regime, and hence the TE estimate is a good approximation
to the actual solution error.}.
The solution obtained on the coarse level interior to this boundary will
not be as accurate; however, the assumed hyperbolic nature of the PDEs
will protect this inaccuracy from polluting the coarse/fine boundary
region within a single coarse level time step.

After $\rhot$ time-steps on level $\ell+1$, when the solution
on grids at levels $\ell$ and $\ell+1$ are again in synchrony, grid
functions from level $\ell+1$ are {\em injected} into the coarse grids
at level $\ell$, in the region of overlap between the two levels. Thus,
the most accurate solution available at a given point $\vec{x}$ is
continuously propagated to all grids in the hierarchy that contain $\vec{x}$.
Injection simply consists of
copying values from level $\ell+1$ to level $\ell$ at common points
(in the more general B\&O algorithm, where finer levels can
be rotated relative to coarser levels, the injection step requires
some form of interpolation).
Fig.~\ref{BO1_ts_fig} is a pseudo code description of the B\&O
time-stepping procedure just described.
\begin{figure}
\epsfxsize=17cm
\begin{flushleft}
\begin{obeylines}
{\tt
~~~~~~repeat $N$ times: call single\_step(1);
~~~~~~stop;
~~~~~~
~~~~~~subroutine single\_step(level $\ell$)
~~~~~~~~~if (regridding time for level $\ell$) then
~~~~~~~~~~~~regrid from levels $\ell$ to $\ell_f$;
~~~~~~~~~end if
~~~~~~
~~~~~~~~~if ($\ell$ >1) then 
~~~~~~~~~~~~set boundary conditions along AMR boundaries at time 
~~~~~~~~~~~~~~~$t_\ell+\Delta t_\ell$ via interpolation from level ($\ell-1$);
~~~~~~~~~end if
~~~~~~~~~
~~~~~~~~~perform 1 evolution step on all grids at level $\ell$;
~~~~~~~~~
~~~~~~~~~$t_\ell:=t_\ell+\Delta t_\ell$;
~~~~~~~~~
~~~~~~~~~if ($\ell<\ell_f$) then
~~~~~~~~~~~~repeat [$\rhot:=\Delta t_\ell/\Delta t_{\ell +1}$] times: call single\_step($\ell+1$);
~~~~~~~~~~~~inject the solution from level $\ell+1$ to level $\ell$
~~~~~~~~~~~~~~~in the region of overlap between levels $\ell$ and $\ell+1$;
~~~~~~~~~end if
~~~~~~~~~
~~~~~~end of subroutine single\_step
}
\end{obeylines}
\end{flushleft}
\caption
{A pseudo-code representation of the Berger and Oliger time stepping
algorithm. 
\label{BO1_ts_fig} }
\end{figure}

\section{Berger and Oliger Style AMR for Constrained Evolution}\label{sec_mod_BO}

The locality argument given in the preceding section
to justify the use of the B\&O evolution algorithm
is only applicable to hyperbolic equations, for then
the finite speed of propagation prevents contamination of the
solution in the boundary region of a coarse level by
a poorly resolved solution in the interior.
Solutions to elliptic equations do not share
this property, and therefore it is not feasible to solve for such equations 
on the coarse grid alone, with the intention
of supplying boundary conditions for subsequent fine grid time 
steps\footnote{Except, as mentioned in the introduction, 
for certain kinds of linear systems, where the source terms appearing
in the elliptic equations can cleanly be identified and coarsened in 
a manner that conserves the source ``energy density''. 
This is not possible for the Einstein equations, which are fundamentally
non-linear, since the gravitational field acts as its own source.}.
One way to circumvent this problem is to abandon the B\&O recursive
time-stepping procedure. In other words, one could evolve the entire hierarchy
forward in time with a global time step, for example 
by performing a Crank-Nicholson style iteration as in the unigrid code
(see Sec.~\ref{sec_unigrid}).
A major drawback to this method is that, to satisfy the CFL condition,
the global time step will need to be set to $\lambda \Delta \rho(\ell_f)$, 
where $\Delta \rho(\ell_f)$ is the cell size on the {\em finest} 
level $\ell_f$ in the hierarchy, and $0 < \lambda \lesssim 1$ is a constant.  
This would require that one take
$\rho_t^{\ell_f-\ell}-1$ additional time steps at 
level $\ell$ for each time step that the usual $B\&O$ algorithm
would have taken.

The technique that we propose here to incorporate elliptic equations 
into the standard B\&O time-stepping framework
is to employ a combination of {\em extrapolation and delayed solution} of the
elliptic variables (see Fig.~\ref{BO_ts_fig2}).
Simply stated, on coarse levels one does {\em not} solve the elliptic equations 
during the evolution step of the algorithm; rather, one extrapolates
the corresponding variables to the advanced time from the solution obtained at
earlier times. The solution of the elliptic equations is delayed until 
{\em after} the injection of fine grid (level $\ell+1$) values into the 
parental coarse grids (level $\ell$).
At that stage, all levels from $\ell$ to $\ell_f$ are in sync, and 
the elliptic equations are solved over the {\em entire resulting subset} 
of the hierarchy, with boundary conditions on AMR boundaries of level $\ell$
set via extrapolation. This ensures that all details 
from finer grids interior to level $\ell$ are represented in the solution.

\begin{figure}
\epsfxsize=17cm
\begin{flushleft}
\begin{obeylines}
{\tt
~~~~~~repeat $N$ times: call single\_step(1);
~~~~~~stop;
~
~~~~~~subroutine single\_step(level $\ell$)
~~~~~~~~~if (regridding time for level $\ell$) then
~~~~~~~~~~~~regrid from levels $\ell$ to $\ell_f$;
~~~~~~~~~end if
~
~~~~~~~~~for hyperbolic variables: if ($\ell > 1$) then set boundary conditions along 
~~~~~~~~~~~~AMR boundaries at time ($t_\ell+\Delta t_\ell$) via interpolation from level $\ell-1$;
~
~~~~~~~~~for elliptic variables: extrapolate the entire grid function to time
~~~~~~~~~~~~($t_\ell+\Delta t_\ell$) from earlier-time values;
~
~~~~~~~~~repeat
~~~~~~~~~~~~if ($\ell=\ell_f$) then perform 1 multigrid vcycle on elliptic variables
~~~~~~~~~~~~~~~at time ($t_\ell+\Delta t_\ell$);
~~~~~~~~~~~~perform 1 iteration of the Crank-Nicholson sweep for all 
~~~~~~~~~~~~~~~hyperbolic variables;
~~~~~~~~~ until (residual norm < tolerance)
~
~~~~~~~~~$t_\ell:=t_\ell+\Delta t_\ell$;
~
~~~~~~~~~if ($\ell<\ell_f$) then
~~~~~~~~~~~~repeat [$\rhot:=\Delta t_\ell/\Delta t_{\ell +1}$] times: call single\_step($\ell+1$);
~~~~~~~~~~~~compute the truncation error estimate for level $\ell+1$ by subtracting 
~~~~~~~~~~~~~~~the solution in the region of overlap between levels $\ell$ and $\ell+1$;
~~~~~~~~~~~~inject the solution from level $\ell+1$ to level $\ell$
~~~~~~~~~~~~~~~in the region of overlap between levels $\ell$ and $\ell+1$;
~~~~~~~~~end if
~
~~~~~~~~~if ($\ell=1$ or ($\ell<\ell_f$ and $t_\ell!=t_{\ell-1}$)) then
~~~~~~~~~~~~$t:=t_\ell$;
~~~~~~~~~~~~do $\ell_0:=\ell+1$ to $\ell_f$
~~~~~~~~~~~~~~~for each elliptic variable $f_{\ell_0}(t)$ : $\hat{f}_{\ell_0}(t):=f_{\ell_0}(t)$;
~~~~~~~~~~~~end do 
~~~~~~~~~~~~re-solve the elliptic equations over the sub-hierarchy [$\ell..\ell_f$] at time $t$;
~~~~~~~~~~~~do $\ell_0:=\ell+1$ to $\ell_f$
~~~~~~~~~~~~~~~for each elliptic variable $f_{\ell_0}(t)$ :
~~~~~~~~~~~~~~~~~~$\Delta f_{\ell_0}(t):=f_{\ell_0}(t)-\hat{f}_{\ell_0}(t)$;
~~~~~~~~~~~~~~~~~~$f_{c\ell_0}(t):=\Delta f_{\ell_0}(t)/\rhot^{\ell_0-\ell}$;
~~~~~~~~~~~~~~~~~~$f_{\ell_0}(t-\rhot\Delta t_{\ell_0}):=f_{\ell_0}(t-\rhot\Delta t_{\ell_0})+\Delta f_{\ell_0}(t)-f_{c\ell_0}(t)$;
~~~~~~~~~~~~end do 
~~~~~~~~~end if
~
~~~~~~end of subroutine single\_step
}
\end{obeylines}
\end{flushleft}
\caption
{A pseudo-code representation of the modified Berger and Oliger time stepping 
algorithm described in Sec.~\ref{sec_mod_BO}; compare to the original
algorithm in Fig.~\ref{BO1_ts_fig}. Notice that here we have expanded
the ``perform 1 evolution step...'' statement in Fig.~\ref{BO1_ts_fig} 
to highlight the fact that in the modified algorithm the finest level
is treated differently than the coarser levels then (though the
{\em particular} details of the evolution step are not significant---for 
concreteness we list the same scheme as used in the unigrid code). Here
we also show where the truncation error estimate is computed when using a self-shadow
hierarchy (see App. \ref{algorithm_details}).}
\label{BO_ts_fig2}
\end{figure} 

\begin{figure}
\begin{center}
\psfrag{s1}{Step 1: Extrapolate prior to CN iteration}
\psfrag{s2}{Step 2: Solve elliptic equations after CN iteration}
\psfrag{s3}{Step 3: Apply correction}
\psfrag{flt}{$f_\ell(t)$}
\psfrag{flht}{$\hat{f}_\ell(t)$}
\psfrag{fcl}{$f_{c\ell}(t)=\frac{{f}_\ell(t)-\hat{f}_\ell(t)}{2}$}
\psfrag{v}{$v=\hat{f}_\ell(t)-f_\ell(t-2\Delta t)$}
\psfrag{fm2dt}{$f_\ell(t-2\Delta t)$}
\psfrag{fm4dt}{$f_\ell(t-4\Delta t)$}
\psfrag{c}{corrected $f_\ell(t-2\Delta t)$}
\psfrag{eq}{$=f_\ell(t)-v-f_{c\ell}(t)$}
\psfrag{t}{$t$}
\psfrag{tmdt}{$t-\Delta t$}
\psfrag{tm2dt}{$t-2\Delta t$}
\psfrag{tm3dt}{$t-3\Delta t$}
\psfrag{tm4dt}{$t-4\Delta t$}
\psfrag{tpdt}{$t+\Delta t$}
\includegraphics[width=14cm,clip=true]{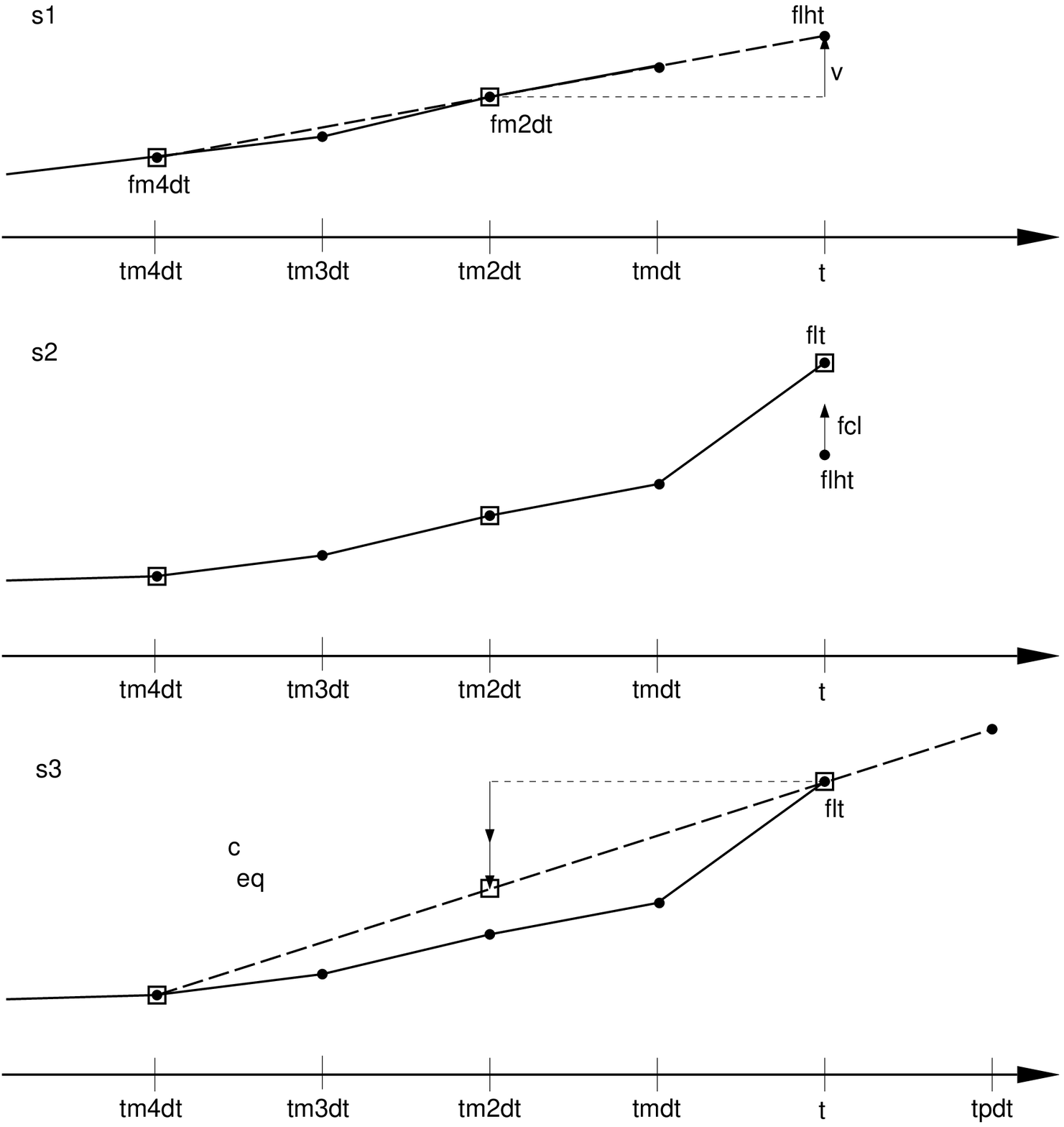}
\end{center}
\caption
{An illustration of the technique used to extrapolate and solve for elliptic 
variables within the AMR framework. In this example, we show the evolution of an
elliptic variable $f_\ell$ from $t-\Delta t$ to $t$, assuming that $\rhot=2$
and $\ell<\ell_f$.  Initially (step 1), $f_\ell$ at 
$t$ is calculated via linear extrapolation from data at past 
times $t-2\Delta t$ and $t-4\Delta t$ (quantities used for extrapolation 
are depicted by boxes in the diagram). This value, labeled
$\hat{f}_\ell(t)$, is used during the solution of the hyperbolic
variables (via Crank-Nicholson iteration (CN) here).
At time $t$, we assume
levels $\ell$ and $\ell_c=\ell-1$ are in sync, and so
after the CN iteration (and {\em after} the equations on finer
levels $\ell+1..\ell_f$ have been evolved to $t$)
the elliptic equations are re-solved over
levels $\ell_c..\ell_f$ (step 2). This results in a change in the 
value of $f$ from $\hat{f}_\ell(t)$ to $f_\ell(t)$ (for simplicity 
we assume that a similar change that occurred at time $t-2\Delta t$ is zero). 
This change is propagated back to
$t-2\Delta t$, so that the same velocity $v$, {\em modulo a correction}
$f_{c\ell}(t)$, will be used to extrapolate $f$ to $t+\Delta t$ (step 3). 
Here, since $\ell-\ell_c=1$, the correction is such that one is effectively
extrapolating from $t-4\Delta t$ and $t$ to time $t+\Delta t$; this would
not be the case otherwise---see (\ref{mg_ext_cor}-\ref{mg_ext_ap_cor}).
(If $\ell=\ell_f$, then the elliptic
variables are solved within the CN iteration, and the value 
obtained afterward is used for $\hat{f}_\ell(t)$.)
\label{mg_extrap_fig}}
\end{figure}

One of the non-trivial aspects of this technique is the method used to
extrapolate the elliptic variables, which we now describe.
We use linear (2nd order) extrapolation in time,
with periodic {\em corrections} to try to account for changes that occur upon 
global multigrid solves. For level $\ell$, $\ell>1$, the two past-times used in the
extrapolation are the two most recent times {\em when levels $\ell$ and $\ell-1$
were in sync} (thus, at times when a solution of the elliptic variables involving
at least levels $\ell-1$ to $\ell_f$ was obtained); in other words, every 
$\rhot$ steps we save the elliptic variables for use in extrapolation\footnote{Early 
experiments using the values from the two most recent time steps of
level $\ell$ for extrapolation resulted in unstable evolution.} (see Sec. \ref{sec_grid_init}
for a pseudo-code description of how the past time levels are initialized
for the very first time step of evolution). For level $\ell=1$, the two most
recent time levels are used for extrapolation\footnote{Though note that in our algorithm level $1$ is {\em always}
fully refined because of the
self-shadow hierarchy mechanism we use for truncation error estimation (see Sec. \ref{sec_ss}),
and therefore level 2 should be
considered the ``true'' coarsest level of the hierarchy (in effect, level 1 is only used
for truncation error estimation)}.
The correction, applied only to levels $\ell>1$, is calculated as follows. Whenever level $\ell$ is in sync 
with level $\ell_c, \ell_c+1,\ldots,\ell_f+1,\ell_f$, where $\ell_c<\ell$, a multigrid solve takes place 
over levels $\ell_c,\ldots,\ell_f$. Denote by $\hat{f}_\ell(t)$ the value
of a variable $f$ at level $\ell$, 
calculated via extrapolation from 
$f(t-\rhot\Delta t_\ell)_\ell$ and $f(t-2\rhot\Delta t_\ell)_\ell$, and let
$f_\ell(t)$ denote the value of the same variable {\em after} the multigrid
solve at time $t$ (note that for simplicity in notation we have dropped the 
spatial coordinate dependence of the variable $f$).
As illustrated in Fig.\ref{mg_extrap_fig}, the correction contains two components:
\begin{equation}\label{mg_ext_cor}
\Delta f_{\ell}(t)\equiv f_\ell(t)-\hat{f}_\ell(t) \ \ \ \mbox{and} \ \ \
f_{c\ell}(t)\equiv\frac{\Delta f_{\ell}(t)}{\rhot^{\ell-\ell_c}}, 
\end{equation}
which are used to change the {\em past} time value $f_\ell(t-\rhot\Delta t_\ell)$ as follows:
\begin{eqnarray}\label{mg_ext_ap_cor}
f_\ell(t-\rhot\Delta t_\ell) \longrightarrow f_\ell(t-\rhot\Delta t_\ell) + \Delta f_{\ell}(t) 
- f_{c\ell}(t)
\end{eqnarray}
The logic behind this form of correction stems from a couple of observations
about how the re-solved solution differs from the extrapolated solution,
and how some adaptive solutions differ from unigrid solutions
of comparable resolution.
First, in general $\Delta f_{\ell}(t)$ is (in a loose sense) proportional
to $\ell_c-\ell$; i.e.~the more levels over which the elliptic equations are
re-solved, the larger the change in the interior, finer level $\ell$ solution.
However, the change in
the interior part of the solution induced by the global solve tends to
be a near constant shift, leaving finer details of the
interior solution unchanged. 
Second, the local ``velocity'' $f_\ell(t)-f(t-\rhot\Delta t_\ell)_\ell$
(calculated {\em prior} to the correction) tends to be represented quite accurately
in the adaptive solution scheme. 
Therefore the  $\Delta f_{\ell}(t)$ part of the correction preserves
this velocity for subsequent extrapolation, and is in fact essential for the
stability of the algorithm with a deep hierarchy, as then the
global shift is often larger in magnitude than the local velocity.
The second part of the correction $f_{c\ell}(t)$ is an attempt to improve the accuracy of the
extrapolation, for if in hindsight the quantity 
$f_{c\ell}(t)$ had been added to $f$ at each one of the 
$\rhot^{\ell-\ell_c}$ intermediate time steps between solves over levels
$\ell_c..\ell_f$, then $\Delta f_{\ell}(t)$ would be zero at time 
$t$ (ignoring, of course, the effect that this putative correction would have 
had on the solution). 

One last comment regarding the extrapolation: on the finest level of 
the hierarchy, the elliptic equations are solved via the usual
interleaved Crank-Nicholson/$V$-cycle iteration as discussed in Sec.~\ref{sec_unigrid};
then the extrapolation simply serves to set boundary conditions.

\section{Results}\label{sec_results}

The extrapolation technique described in the previous section
is rather ad-hoc, though both the choice of 
which past times to extrapolate from and the use of corrections play a 
significant role in the stability and accuracy of the adaptive evolution code. In
this section we present some results showing the effectiveness
of the algorithm.

\subsection{Comparison with a Unigrid Evolution}\label{sec_res_comp}

The first test of the algorithm presented here is
a comparison of unigrid evolutions
to a similar AMR evolution (test results on the convergence properties
of the unigrid code can be found in \cite{paper1}). 
Specifically, we compare an evolution obtained with the AMR code
to a unigrid evolution, where the entire unigrid mesh is given the
resolution $h$ of the finest level in the AMR hierarchy. To gauge how
well the AMR solution approximates the unigrid one, we then compare the
unigrid solution to that obtained with two additional unigrid runs, with
resolutions of $2h$ and $4h$.
In certain respects this is not a very stringent test, as limited computational resources
do not allow us to run using very high resolution unigrid simulations (which
of course is the motivation for pursuing AMR). Also, the accuracy
of the AMR solution will largely depend on the structure of the grid hierarchy,
which is (predominantly) controlled by the maximum allowed TE. Thus, in
principle one could obtain better agreement between the AMR and unigrid
simulations by decreasing the TE parameter while keeping the maximum
depth of the AMR hierarchy fixed. Nevertheless, this comparison
{\em does} demonstrate that the adaptive algorithm works in practice.

The initial data for this example is a time symmetric 
scalar field pulse:
\begin{equation}\label{sf_id}
\Phi(0,\rho,z)=  A \exp \left[ -\frac{\rho^2 + z^2}{\Delta^2} \right],
\end{equation}
with $A=0.23$ and $\Delta=1.0$ (all other free fields are set to $0$ at $t=0$).
This amplitude is sufficiently large that the solution is in the non-linear regime
(and thus close to forming a black hole).
The outer boundary is at $\rho_{\rm max}=z_{\rm max}=-z_{\rm min}=10$. 
The maximum value for the TE is set to $10^{-5}$ 
(only $\Phi$ and $\Pi$ are used in the calculation of the TE---see App. \ref{algorithm_details}).
The resolution of the base grid is $65\times129$, with
$\rhosp=2$, and up to $3$ additional levels of refinement. Thus, level $4$ 
has an effective resolution of $513\times1025$, and so we choose 
unigrid comparison runs with resolutions of
$513\times1025$ ($h$), $257\times513$ ($2h$) and $129\times257$ ($4h$).
The Courant factor is $\lambda=0.3$ for all runs, and the physical
time at the end of each simulation is 
$t\sim2.8$ (which corresponds to 480 time steps on the level with the finest resolution).

\begin{figure}
\begin{center}
\psfrag{rho}{$\rho$}
\psfrag{z}{$z$}
\includegraphics[width=14cm,clip=true]{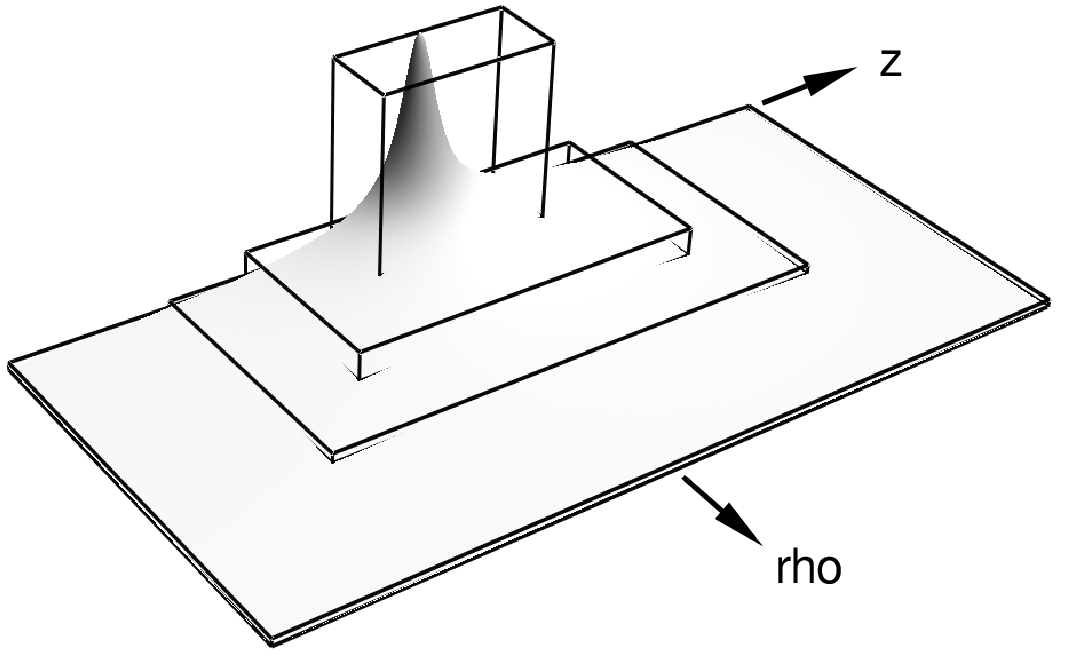}
\end{center}
\caption
{A surface plot of $\psi$ at $t=2$ from the adaptive simulation discussed
in Sec.~\ref{sec_res_comp}, where the
height of the surface is proportional to the magnitude of $\psi$ (ranging
approximately from $1$ at the outer boundary to $1.5$ at the origin $\rho=z=0$). 
Overlaid on the surface are the AMR grid bounding boxes---the smallest, interior
box has the finest resolution, corresponding to an effective unigrid resolution
of $513\times1025$.
\label{fig_psi_ad_t2}}
\end{figure}

\begin{figure}
\begin{center}
\includegraphics[width=16cm,clip=true]{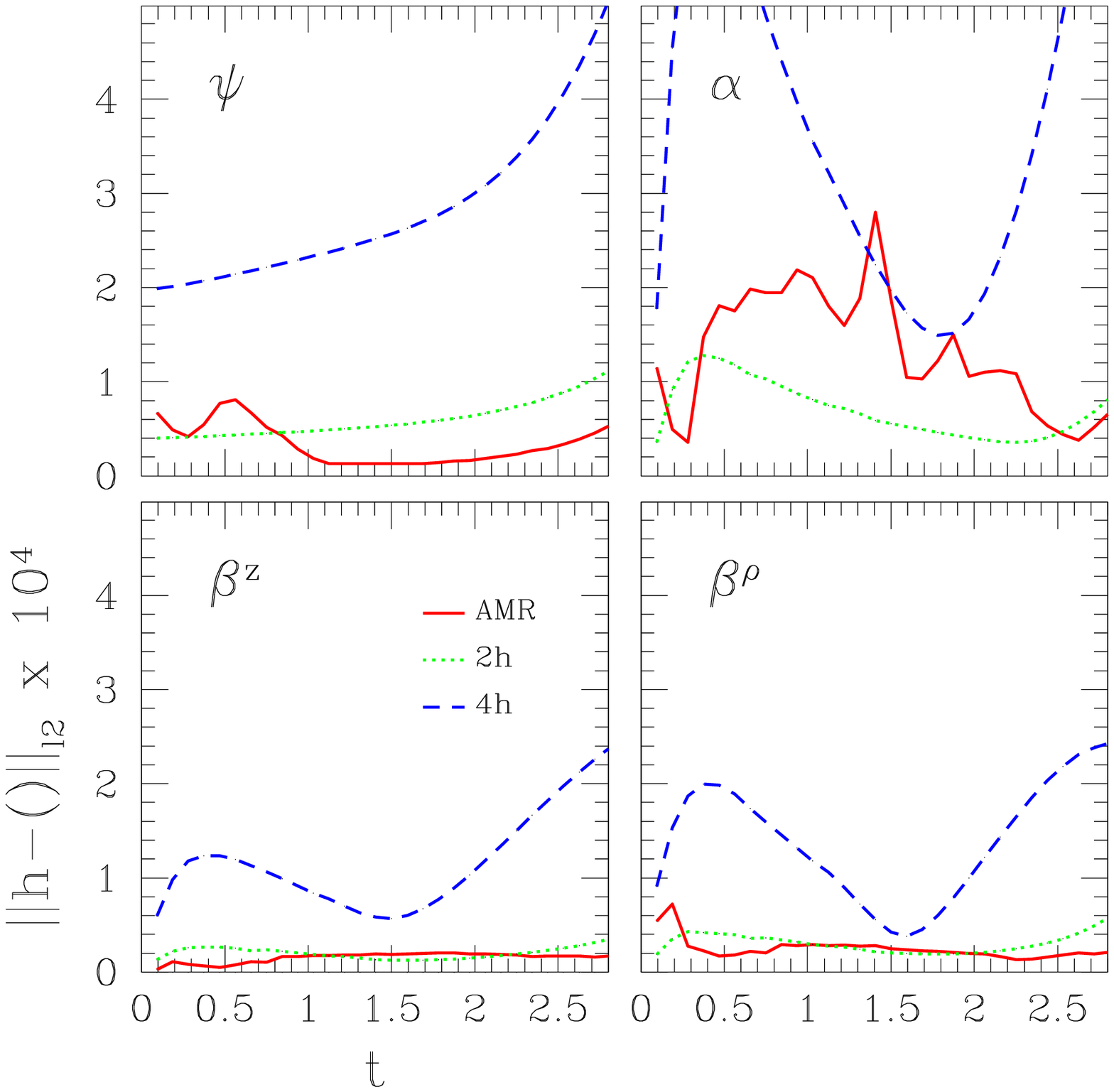}
\end{center}
\caption
{Comparison between unigrid and adaptive simulation results for
the elliptic variables $\psi,\alpha,\brho$ and $\bz$, as discussed in 
Sec.~\ref{sec_res_comp}. Shown here are $\ell_2$ norms of differences
between the solution generated by the $h$ resolution ($513 \times 1025$) 
unigrid simulation and each of the solutions produced by two lower resolution 
unigrid simulations, $2h$ ($257\times513$) and $4h$ ($129\times257$), and an
adaptive simulation (see Fig.~\ref{fig_psi_ad_t2} for a representative
sample of the mesh structure from the AMR solution at $t=2$, where the 
base level $1$ has resolution $65\times129$, and the finest level $4$ has the 
same resolution as the $h$ unigrid run). Thus, we are using the $h$ unigrid solution as
a benchmark, and the plots show that the adaptive solution is of
comparable accuracy to the $2h$ unigrid solution, and of significantly
greater accuracy than the $4h$ unigrid solution, despite that the majority
of the coordinate domain of the AMR solution is covered by grids
with mesh spacing $h_l$ satisfying $h_l > 4h > 2h$.
\label{fig_l2_uni_comp}}
\end{figure}

\begin{figure}
\begin{center}
\includegraphics[width=16cm,clip=true]{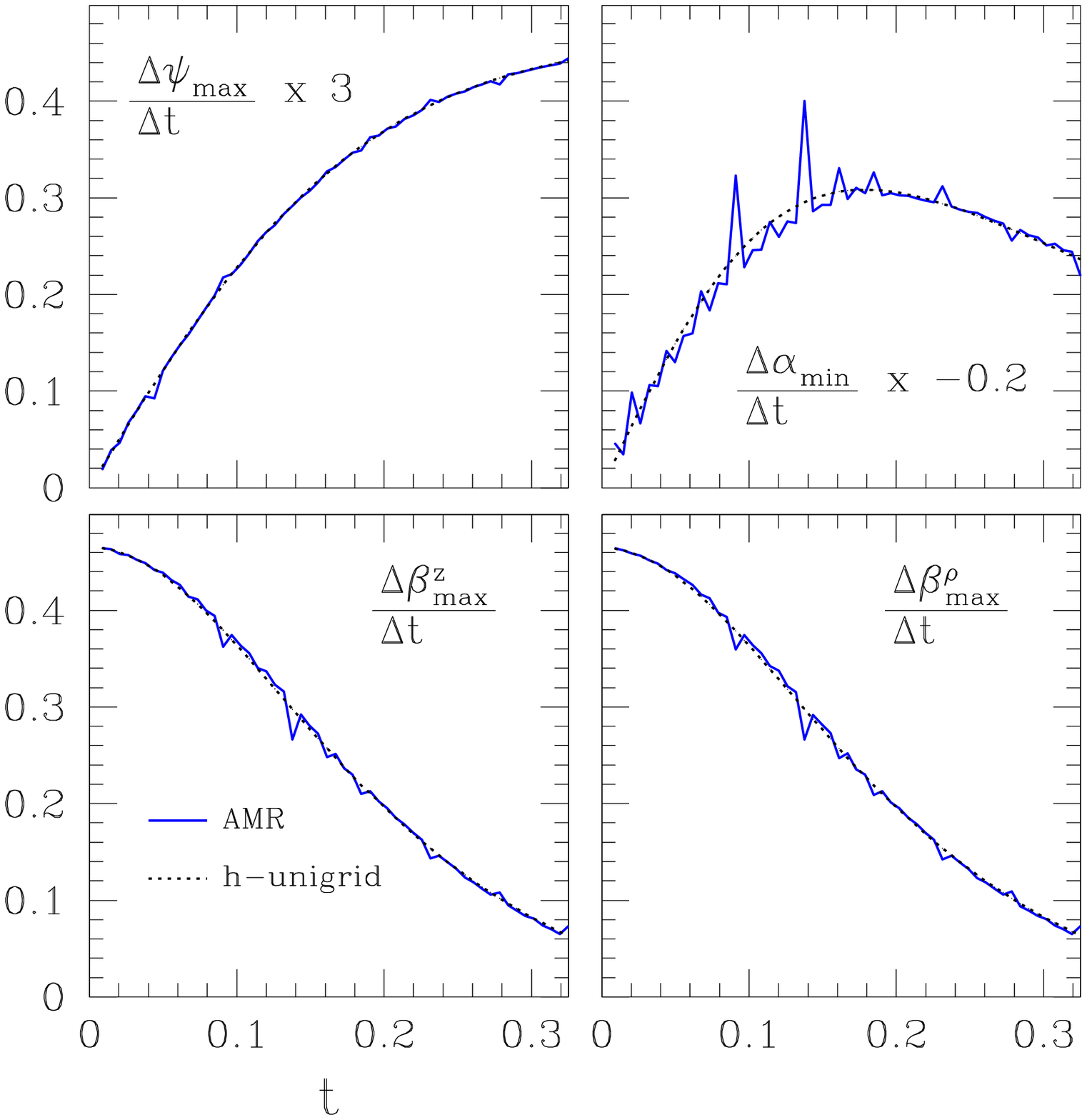}
\end{center}
\caption
{Comparison between $h$ resolution ($513\times1025$) unigrid and adaptive 
simulations results as discussed in Sec.~\ref{sec_res_comp}, 
showing the time-difference of the maximum (or minimum, as appropriate) of the 
elliptic variables $\psi,\alpha,\brho$ and $\bz$. For clarity, only about
$50$ time-steps are shown, and the differences for $\alpha$ and $\psi$ have been
scaled by constants to fit all the plots on the same vertical scale. The 
time-difference $\Delta f/\Delta t$ is calculated as $(f^{n+1}-f^{n})/\Delta t$, 
where $f^{n}$ labels one of the above quantities at time step $n$. These 
figures demonstrate a couple of interesting aspects of the extrapolation
scheme for the adaptive code. First, the high frequency ``noise'' that is apparent 
in the adaptive solution (which has been exaggerated here by taking 
a time-difference) does not adversely affect the accuracy of the solution
on average. Second, the presence of such noise suggests why linear extrapolation
from the most recent time levels is unstable (though it is not obvious why
extrapolation from past time levels that are in-sync with a parent level results
in stable evolution).
\label{fig_ddt_comp}}
\end{figure}

Fig.~\ref{fig_psi_ad_t2} shows a plot of the conformal factor $\psi$ 
at $t=2$ from the adaptive simulation, with grid bounding boxes overlaid
to give an idea of the structure of the hierarchy adaptively generated by the AMR code.
Fig.~\ref{fig_l2_uni_comp} shows $\ell_2$ norms of the differences
between the solutions generated by the $h$ resolution unigrid 
simulation and the two lower resolution unigrid and adaptive simulations.
For brevity we only show differences for the $4$ elliptic variables;
differences in the hyperbolic variables exhibit similar behavior.
What Fig.~\ref{fig_l2_uni_comp} demonstrates is that, compared to
the $h$ unigrid simulation, the adaptive solution has accuracy comparable to
the $2h$ simulation, and significantly greater
accuracy than the $4h$ simulation, even though (as illustrated in 
Fig.~\ref{fig_psi_ad_t2}) the majority of the coordinate domain in the
adaptive solution is spanned by a grid with less resolution that either
that of the $2h$ or $4h$ simulations.
Fig.~\ref{fig_ddt_comp} is a comparison of the time-difference of the maximum 
(or minimum, as appropriate) of the elliptic variables $\psi,\alpha,\brho$ and 
$\bz$ from the $h$ unigrid and adaptive simulations.
The time-difference $\Delta f/\Delta t$ is calculated as $(f^{n+1}-f^{n})/\Delta t$,
where $f^{n}$ denotes one of the quantities, $\psi, \alpha, \brho$ or $\bz$ at time step $n$. 
Fig.~\ref{fig_ddt_comp} demonstrates two interesting aspects of the extrapolation
scheme of the adaptive code. First, the high frequency temporal ``noise'' that is apparent
in the adaptive solution of the elliptic variables (and that
has been exaggerated in the figure by taking
a time-difference) does not adversely affect the accuracy of the solution
on average. Second, the presence of such noise suggests 
why linear extrapolation
at level $\ell$ from the most recent times of level $\ell$ is unstable; however, it 
is not so obvious why extrapolation from past times that are in-sync with a parent 
level ($\ell-1$ or less) results in stable evolution. Note that with the particular
system of equations solved for here (summarized in App. \ref{appendix})
{\em no} explicit time derivatives of any elliptic quantities appear. Furthermore,
during the evolution phase of the algorithm (see Sec.~\ref{sec_unigrid}),
the Crank-Nicholson differencing scheme only couples the time average (over two
time steps) of elliptics variables to the hyperbolic equations, providing
a certain amount of temporal smoothing. Thus, the results shown in 
Fig.~\ref{fig_ddt_comp} suggest that modifications to the extrapolation
scheme may be needed for systems of PDEs that use other methods to difference
in time, or if time derivatives of variables solved for using elliptic
equations couple to the hyperbolic equations.

\subsubsection{Timing information}\label{sec_timing}

We conclude this section by presenting timing information in Tab. \ref{tab_time} below 
for the comparison test just described. This serves to show that,
at least for the elliptic-hyperbolic system considered here,
the overhead of the AMR algorithm is negligible, and that the 
solution time scales roughly linearly with the total number of grid points in the 
discrete solution.
See also Sec.~\ref{sec_timing2}, which contains more detailed information on 
the percentage of time spent in various stages of the algorithm for the 
test simulations presented in the next section.

\begin{table}
\begin{tabular}[t]{ | c || c | c | c || c | }
\hline
 & unigrid-4h & unigrid-2h & unigrid-h & AMR \\
\hline
\hline
runtime ($s$)  & $4.45 \times 10^2$ & $4.48 \times 10^3$ & $3.31 \times 10^4$ & $2.19 \times 10^3$ \\
total number of grid points & $4.01 \times 10^6$ & $3.18 \times 10^7$ & $2.54 \times 10^8$ & $1.97 \times 10^7$ \\
average time per grid point ($\mu s$) & $111$ & $141$ & $131$ & $111$ \\
\hline
\end{tabular}
\caption{Timing information for the tests described in Sec.\ref{sec_res_comp}.
The {\em runtime} is the wall time taken (on a 2.4Ghz Intel XEON processor), in seconds, for each simulation,
including initial data calculation and evolution. The {\em total number of grid points}
is a count of all the grid points, in space and time, at which a solution was obtained 
during the simulation. This includes the grid points used during calculation
of the initial data, and the calculation of the initial hierarchy for the adaptive run.
The {\em average time per grid point} (in microseconds) is the runtime divided by the total number of grid points.
What these numbers suggest (see also Tab. \ref{tab_time2} is that the 
computational cost of this solution method
scales approximately linearly with the total number of grid points, and that the 
computational overhead for the adaptive algorithm is negligible. 
Note however that we have {\em not} taken
into account that in the adaptive hierarchy certain coordinate locations are covered
by multiple points, and therefore there is some ``wastage'' in the AMR calculation.
In this particular simulation, at any one time between $15-20\%$ of the points in 
the adaptive hierarchy were redundant, and this should be viewed as an additional
(and unavoidable) overhead to the per-point cost of this Berger and Oliger style AMR algorithm.
}
\label{tab_time}
\end{table}

\subsection{Convergence and Consistency Tests}\label{sec_ad_cvt}

In this section we present some convergence test results, providing evidence
that the overall AMR solution scheme {\em is} stable and convergent. 
We also compare the solution
of the conformal factor $\psi$ to that obtained in a {\em partially
constrained} evolution, where $\alpha,\brho$ and $\bz$ are solved as described in
Sec.~\ref{sec_graxi} via the slicing condition and momentum constraints respectively,
but where a hyperbolic evolution equation (\ref{psi_evo_eqn}) obtained from the maximal slicing 
condition $K=0$, rather that the elliptic Hamiltonian constraint, is now used to 
update $\psi$ \footnote{That an independent equation (other than the 
Hamiltonian constraint) exists for $\psi$ is due to the over-determinism
in the Einstein equations, as discussed in the introduction.}. In the 
continuum limit, and the limit where the outer boundary position goes
to infinity, the solution obtained from
fully and partially constrained evolution should (assuming both
numerical implementations are consistent and stable) converge to a unique
solution of the Einstein equations. In the partially constrained evolution,
$\psi$ is evolved as a hyperbolic variable, and thus within the
traditional B\&O time stepping framework; however, during fully constrained
evolution $\psi$ is solved for using an elliptic equation using 
the new AMR technique. Thus, demonstrating convergence to a consistent solution
for $\psi$ is a rather non-trivial test of the modified AMR algorithm.

We use the following technique to calculate convergence factors
for the adaptive code. We choose a ``modest'' value ($10^{-3}$ in this case)
for the maximum allowed TE (calculated as described in Sec. \ref{sec_ss}), 
run a simulation, and save a copy of the dynamical
grid structure produced during the evolution. The solution on this
grid hierarchy will serve as the coarsest resolution simulation,
labeled $4h$. For the higher resolution simulations $2h$ and $h$,
We rerun the code with identical initial data, and use the
same grid hierarchy produced by the $4h$ case {\em except} that the resolution 
of all grids in the hierarchy is doubled (quadrupled) for the $2h$ ($h$)
simulation; the Courant factor is kept constant, hence the
number of time-steps taken per level also doubles (quadruples).
Then, by assuming the usual Richardson expansion, one can compute
a convergence factor $Q^h_f$ for a variable $f$ via
\begin{equation}\label{Q_def}
Q^h_f \equiv \frac{ || f_{4h} - f_{2h} ||_2 }
                  { || f_{2h} - f_{h}  ||_2 },
\end{equation}
where $f_{4h}$ is the solution on the $4h$ hierarchy, and similarly
for $f_{2h}$ and $f_{h}$. In (\ref{Q_def}), the subtraction of grid
functions is only defined at common points between the hierarchies
(every other point of the finer mesh in this case).
The $\ell_2$ norm of a grid function $f$ is taken point-wise as follows
\begin{equation}
||f||_2 = \frac{\left(\displaystyle\sum_{\ell}\displaystyle\sum_{g}\displaystyle\sum_{i,j}
f_{\ell}(\rho_g+i\Delta\rho_\ell,z_g+j\Delta z_\ell)^2\right)^{1/2}}
{\displaystyle\sum_{\ell}\displaystyle\sum_{g}\displaystyle\sum_{i,j}1},
\end{equation}
where the sum over $\ell$ is the sum over all levels in the hierarchy
where $f$ is defined (so for the differences in (\ref{Q_def}) this will
be all levels except the finest), the sum over $g$ is over all grids at the given 
level, and the sum over $(i,j)$ covers all points within a given grid, where $(\rho_g,z_g)$
is the location of the $(i=0,j=0)$ point.
Note that such a point-wise norm is biased (compared to an area-weighted norm, 
such as an integral norm) toward the highest resolution
region of the domain, where most of the points are clustered. This is
desirable for the solution presented here, where the region of
high refinement is centered on a very small part of the domain, and
outside of this region the solution is slowly varying and well represented
by the coarse mesh. An area-weighted norm in this case would almost completely
ignore information on the finest levels. 
For a ``residual'' function $R$, in other words
a function that should converge to zero in the limit as $h\rightarrow 0$, 
we use the following expression for its convergence factor:
\begin{equation}\label{Q_R_def}
Q^h_R \equiv \frac{ || R_{2h} ||_2 }
                  { || R_{h}  ||_2 }.
\end{equation}
For a second-order accurate finite difference scheme
one expects both $Q^h_f$ and $Q^h_R$ to approach $4$ asymptotically. 

The initial data for this example is also a time symmetric 
scalar field pulse given by (\ref{sf_id}), with
$A=0.25$ and $\Delta=0.5$, and all other free fields set to $0$ at $t=0$.
Adopting apparent horizon detection as 
our operative definition of black hole existence,
in this case a black hole (with mass $M\sim 0.12$ in geometric units) 
{\em does} form by $t\sim2.5$.  

\begin{figure}
\begin{center}
\includegraphics[width=16cm,clip=true]{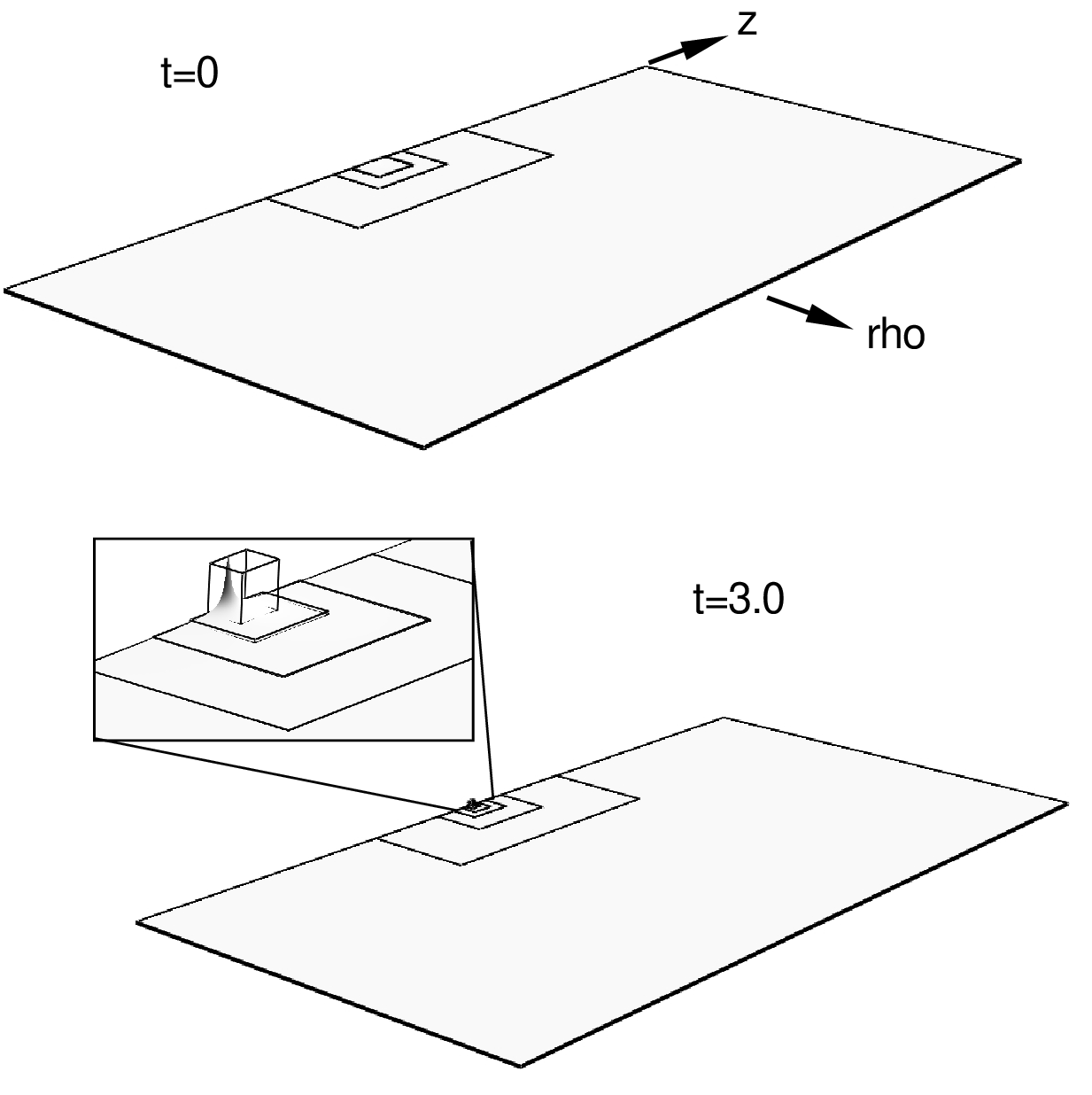}
\end{center}
\caption
{Surface plots of $\psi$ at $t=0$ and $t=3$ from the $h$ adaptive simulation 
discussed in Sec.~\ref{sec_ad_cvt}, where the
height of each surface is proportional to the magnitude of $\psi$ (ranging
approximately from $1.00$ ($1.00$) at the outer boundary to $1.24$ ($3.20$) 
at the origin $\rho=z=0$ at $t=0$ ($t=3$)).
Overlaid on the surfaces are the AMR grid bounding boxes---there are 
4 levels of $2:1$ refinement at $t=0$, 7 levels at $t=3$, and the 
base (coarsest) level has a resolution of $257\times513$.
\label{fig_psi_ad_nts4}}
\end{figure}

\begin{figure}
\begin{center}
\includegraphics[width=16cm,clip=true]{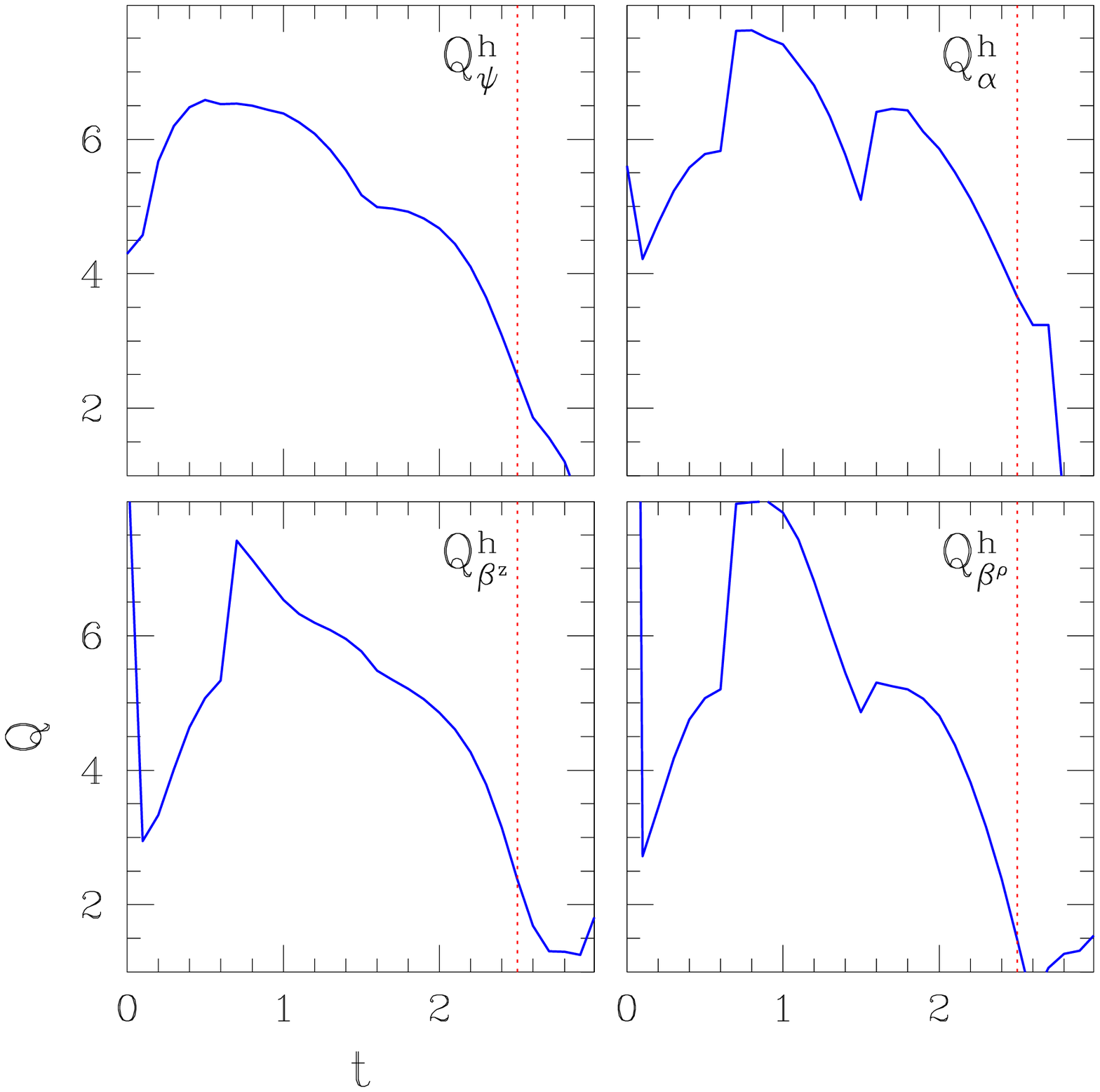}
\end{center}
\caption
{Convergence factors (\ref{Q_def}) for the variables $\psi,\alpha,\brho$
and $\bz$, calculated from the fully constrained adaptive simulations
discussed in Sec.~\ref{sec_ad_cvt}. For second-order accurate finite differencing 
one would expect $Q\approx4$. Time symmetric initial data was used in this case,
for which the exact solution for $\brho$ and $\bz$ at $t=0$ is $0$, hence the
anomalous spikes for the corresponding convergence factors then
(i.e.~$Q\sim0/0$). A black hole
forms at $t\sim2.5$ (denoted by the dashed vertical line), after which significant gradients in metric functions
develop due to the ``grid-stretching'' property of maximal slicing as the spacetime
singularity is first approached, but ultimately avoided; this appears to be the dominant factor causing these
3 simulations ($h$, $2h$ and $4h$) to start to depart from the convergent 
regime (though for any given resolution one expects departure from convergence
to eventually occur).
We cannot explain why the convergence factor is somewhat greater than
four during intermediate times in the simulation, however this is {\em not}
atypical behavior in simulations we have looked at.
\label{fig_l2_adnts4}}
\end{figure}

\begin{figure}
\begin{center}
\includegraphics[width=12cm,clip=true]{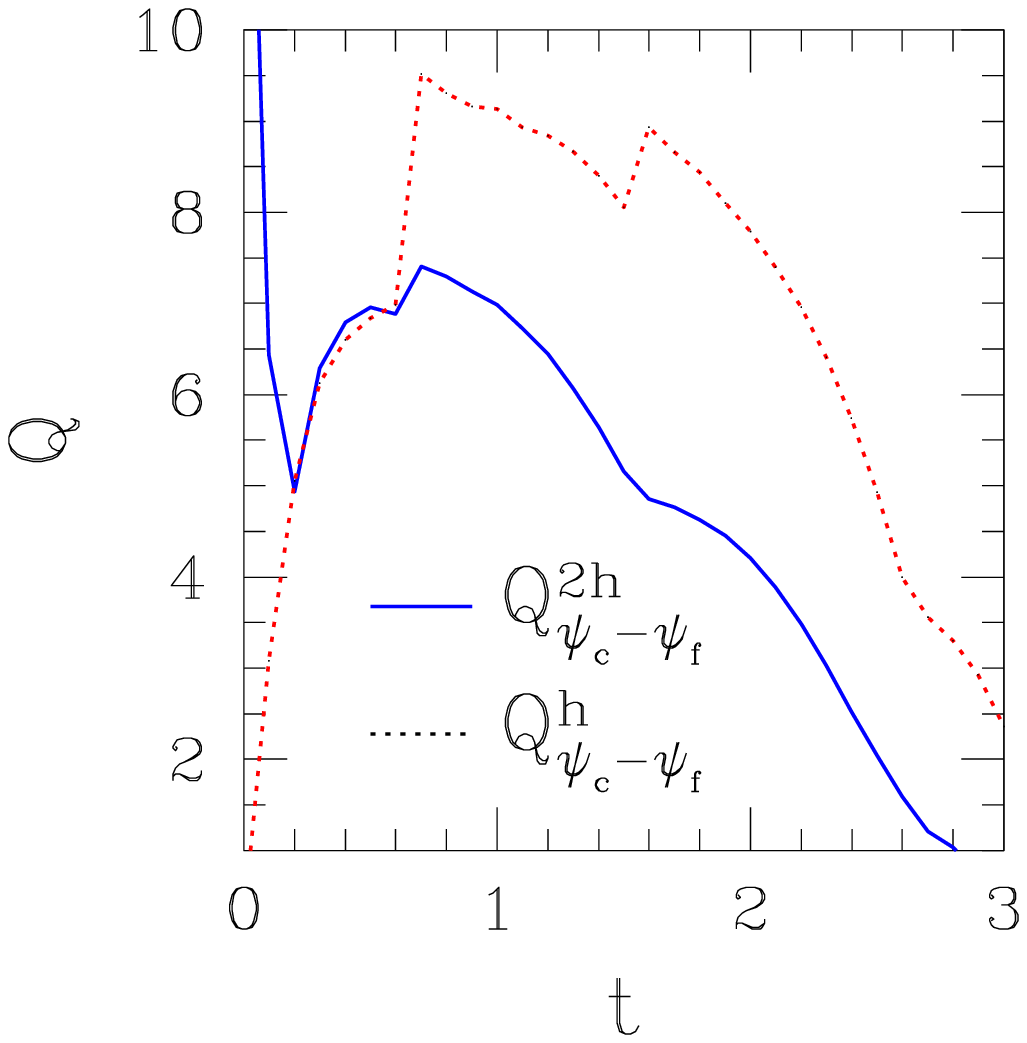}
\end{center}
\caption
{Convergence factors (\ref{Q_R_def}) for the assumed residual quantity $\psi_c-\psi_f$
from the adaptive simulations discussed in Sec.~\ref{sec_ad_cvt},
where $\psi_c$ is the conformal factor $\psi$ from the fully constrained simulation,
and $\psi_f$ is the conformal factor obtained via free evolution. At $t=0$ both
$\psi_f$ and $\psi_c$ are calculated by solving the Hamiltonian constraint, hence
the anomalous behavior in the convergence factor then. Moreover, the boundaries
conditions employed are only consistent with the full set of Einstein equations 
in the limit where the outer boundary position goes to infinity \cite{paper1}.
Early on in the simulation this inconsistency produces a difference in 
$\psi_c-\psi_f$ that is of the same magnitude as the truncation error of the
$h$ simulation, and this causes $Q^h_{\psi_c-\psi_f}$ to start below $4$ near $t=0$. 
However, as evolution proceeds
and gravitational collapse occurs, 
the gradients in $\psi$ in the central region grow rapidly,
and the truncation error component 
of $Q^h_{\psi_c-\psi_f}$ begins to dominate.
As with the results shown in Fig.~\ref{fig_l2_adnts4}, grid-stretching effects
apparently cause the decrease in $Q$ after the black hole forms near $t\sim2.5$.
However, in this plot we can see a trend to improved convergence results
at any given time when resolution is increased from $Q^{2h}$ to $Q^h$.
\label{fig_l2_dpsi}}
\end{figure}

Computationally, relevant parameter 
settings are as follows.
The outer boundary is at $\rho_{\rm max}=z_{\rm max}=-z_{\rm min}=32$,
and for the $4h$ simulation the resolution of the base grid is $65\times129$.
The maximum value for the TE is set to $10^{-3}$; this results in
a grid hierarchy containing 3 additional levels ($\rhosp=2:1$) at $t=0$, and
$6$ additional levels at the end of the 
simulation\footnote{the increase in hierarchy depth that
occurs when black holes form is partly due to the ``grid-stretching'' phenomena 
associated with maximal slicing.}. 
See Fig.~\ref{fig_psi_ad_nts4}
for sample plots of $\psi$ at $t=0$ and $t=3$ to illustrate the grid hierarchy. 
The effective finest grid resolution for the $h$ simulation is 
roughly $16,000$ x $32,000$,
making it impractical to do a unigrid comparison in this case.
Fig.~\ref{fig_l2_adnts4} shows the calculated convergence factors for the
four elliptic quantities $\psi,\alpha,\brho$ and $\bz$, and Fig.~\ref{fig_l2_dpsi}
contains the convergence factor for $\psi_c-\psi_f$, where $\psi_c$ is the conformal
factor $\psi$ from fully constrained evolution, and $\psi_f$ is the conformal
factor calculated from the free evolution of $\psi$. These plots show reasonable
convergence and consistency results, with a couple of caveats discussed 
in the captions.

\subsubsection{Timing information}\label{sec_timing2}

Tab. \ref{tab_time2} contains some timing information for the set of simulations
described in this section. The major point we would like to emphasize with
this table (see the caption for more information) is that solution of the
elliptic equations is {\em not} significantly more expensive than the solution
of the hyperbolic equations in our model; moreover the scaling of the solution time
with problem size is close to linear, again for {\em both} types of equations. 
This is not surprising or new in any sense, as one can {\em immediately}
predict this ``Golden Rule'' scaling behavior from Brandt's work on multi-level adaptive 
(MLAT) schemes \cite{brandt}; indeed, that elliptic equations could be solved
in linear time within the context of general relativity was already demonstrated 
in the early 1980's \cite{MWC_msc,MWC_86}.

\begin{table}
\begin{tabular}[t]{ | c || c | c | c || }
\hline
 & AMR-4h & AMR-2h & AMR-h  \\
\hline
\hline
runtime ($s$)                                   & $3.38 \times 10^2$ & $2.00 \times 10^3$ & $1.40 \times 10^4$ \\
\hline
percentage of runtime solving elliptics equations  & $68.2\%$ & $60.3\%$ & $50.8\%$ \\
percentage of runtime solving hyperbolic equations & $24.4\%$ & $33.1\%$ & $43.1\%$ \\
percentage of runtime in AMR related functions     & $ 3.4\%$ & $ 3.4\%$ & $ 3.8\%$ \\
percentage of runtime in miscellaneous functions   & $ 4.0\%$ & $ 3.2\%$ & $ 2.3\%$ \\
\hline
\end{tabular}
\caption{Timing information for the fully constrained simulations 
described in Sec.\ref{sec_ad_cvt}.
The {\em runtime} is the wall time taken (on a 2.4Ghz Intel XEON processor), in seconds, for each simulation, including initial data calculation and evolution. For this
set of simulations, the total number of grid points in space and time
(compare Tab. \ref{tab_time}) increases by a factor of 8 from the $4h$
to $2h$ simulation, and again by a factor of $8$ from the $2h$ to $h$ simulation. Hence
the runtime should increase by a corresponding factor for an algorithm whose cost
scales linearly with the number of points---we do, approximately, see this behavior.
The next four rows give a rough breakdown of the percentage of total time
spent in each of the key parts of the program (calculated using the function profiling
option of Portland Group's {\em pg} compilers). The two main points we
want to stress with this data is that the scaling (with problem size) of the 
adaptive multigrid algorithm used to solve the elliptic equations is
close to {\em linear}, and is {\em not} significantly slower than the solution of the
hyperbolic equations. Second, the cost of AMR related functions
(including regridding, truncation error estimate calculation, interpolation 
and injection functions) is quite small compared to the cost of 
solving the numerical equations. The miscellaneous functions of the last row
include calculations of diagnostic quantities, the cost of the apparent horizon
finder (that searches for the presence of black holes), and I/O.
}
\label{tab_time2}
\end{table}

\section{Conclusions}\label{sec_conclusion}

In this paper we have discussed modifications of the standard Berger and
Oliger adaptive mesh refinement method, so that the resulting algorithm can 
solve systems resulting from the discretization 
of {\em coupled, non-linear, hyperbolic and elliptic equations} in {\em linear time}.
Moreover, as we have retained recursive time-stepping, the algorithm
is still optimally efficient in solving systems that contain 
hyperbolic equations which are subject to a CFL stability condition. 
The initial application that drove development of the method was a study
of critical gravitational collapse of the scalar field 
in axisymmetry \cite{paper1,paper2,paper3}.  This 
involved the approximate solution of a mixed elliptic/hyperbolic system: the coupled 
Einstein-Klein Gordon system in a certain coordinate system,
with a particular choice among the overdetermined set of PDEs at our disposal to 
advance the solution in time. However, the algorithm is sufficiently general that it 
can be applied to a variety of similar systems of partial differential equations. 
For example, there are numerous problems in astrophysics that need
to evolve various matter equations coupled to gravity and/or the
electromagnetic field, including cosmological structure formation, 
stellar evolution, supernovae, jets, accretion disks, etc. Many of
these scenarios have a large range of spatio-temporal length scales
that need to be modeled,
and can benefit tremendously from Berger and Oliger style AMR. In some 
situations Newtonian gravity is sufficient to accurately describe
the physics, and due to the linearity of the Poisson equation,
modifications to Berger and Oliger as described here are not strictly
necessary. However, some of the most interesting astrophysical events
occur in regions where gravity is sufficiently strong that nonlinear
effects become important, and the algorithm described in this paper could
be of significant use in such simulations\footnote{We note that
even the conformally flat approximation to the field equations, 
which seems to be an adequate extension of Newtonian gravity for 
a certain class of problems \cite{cook_et_al,dimmel_et_al},
requires solution of a non-linear elliptic equation.}.

With regards to new applications of this algorithm in numerical relativity, of 
particular interest is constrained evolution in 3D,
which, on the basis of substantial 1D and 2D 
evidence\cite{bardeen_piran,stark_piran,evans1,nakamura,evans2,
shapiro_teukolsky,abrahams_evans2,choptuik,
abrahams_evans,abrahams_et_al,choptuik_et_al,liebling_choptuik,
garfinkle_duncan2,olabarietta_choptuik,paper1,paper2,ventrella_choptuik,paper3},
we have long felt has great potential for the study of problems such as
black hole collisions, critical gravitational collapse, and the structure of 
black hole interiors. All of the aforementioned lower dimensional studies
made use of the symmetries in the problem, in addition to particular
coordinate choices, to obtain well-posed coupled
elliptic/hyperbolic systems, and so there is some skepticism
in the community about whether constrained evolution can be implemented for
general problems in 3D. Though recently a fully constrained 3D evolution 
scheme was proposed 
in \cite{bonazzola_et_al}, based on the Dirac gauge and spherical
coordinates (the 
implementation presented in \cite{bonazzola_et_al} made use of a 
multidomain spectral solution method). A potential disadvantage
of this system is that it is not obvious how to generalize it 
beyond spherical coordinates, which are not well adapted to studying
problems that are far from spherical symmetry. A formulation of the Einstein equations
that is amenable to 3D constrained evolution and {\em can} be written in 
Cartesian coordinates was described in \cite{andersson_moncrief}.
Both of these formulations are prime candidates for numerical implementation
using our AMR algorithm.

\section{Acknowledgments}
FP would like to thank Bernd Br\"{u}gmann for helpful suggestions on an
earlier version of the manuscript.
The authors gratefully acknowledge research support from
CIAR, NSERC, NSF PHY-0099568, NSF PHY-0244906,
and Caltech's Richard Chase Tolman Fund.
The simulations described here were performed on UBC's {\bf vn}
cluster (supported by CFI and BCKDF), and the {\bf MACI} cluster
at the University of Calgary (supported by CFI and ASRA).
\appendix


\begin{appendix}


\section{Equations and Finite Difference Operators}\label{appendix}

Here, for completeness, we list all of the equations introduced
in Section \ref{sec_graxi}, and the specific set of finite difference
operators used to discretize them.

\subsection{Equations}\label{sec_analytic}
Summarizing Section \ref{sec_graxi}, the four-metric is
\begin{eqnarray}
ds^2 = \left(-\alpha^2+\psi^4\left[(\brho)^2+(\bz)^2\right]\right) dt^2
+ 2\psi^4\left(\brho d\rho + \bz dz\right) dt \nonumber\\
+ \psi^4\left(d\rho^2 + dz^2 + \rho^2 e^{2\rho\sigmabar} d\phi^2\right)
\end{eqnarray}
The conjugate variable to the scalar field $\Phi$ is $\Pi$ (\ref{phi_eqn}),
and the conjugate to $\sigmabar$ is $\omegabar$
(\ref{omegabar_def}). All of these variables are functions of $\rho,z$
and $t$.
The maximal slicing condition results in the following
elliptic equation for $\alpha$:
\begin{eqnarray}
2 \left( \rho \alpha_{,\rho} \right)_{,\rho^2} + \alpha_{,zz}
        + \alpha_{,\rho} \left(  2 \frac{ \psi_{,\rho} }{ \psi }
                               + \left( \rho \sigmabar \right)_{,\rho}
                               \right)
        + \alpha_{,z   } \left(  2 \frac{ \psi_{,z   } }{ \psi }
                               + \left( \rho \sigmabar \right)_{,z   }
                               \right)
\nonumber\\
   - \frac{\psi^4}{2\alpha}
       \left[  (  \beta^\rho{}_{,\rho} - \beta^z{}_{,z}      )^2
             + (  \beta^\rho{}_{,z   } + \beta^z{}_{,\rho} )^2 \right]
   - \frac{\psi^4}{6\alpha} \left[
                                    2 \alpha \rho \omegabar
                                  + \beta^\rho{}_{,\rho}
                                  - \beta^z{}_{,z}
                                 \right]^2
   - 16 \pi \alpha \Pi^2 = 0
\label{slicing_eqn}
\end{eqnarray}
The Hamiltonian constraint is
\begin{eqnarray}
          8 \frac{ \psi_{,\rho \rho} }{ \psi}
       +  8 \frac{ \psi_{,z    z   } }{ \psi}
       + 16 \frac{ \psi_{,\rho^2} }{ \psi }
       +  8 \left( \rho \sigmabar \right)_{,\rho}
              \frac{ \psi_{,\rho  } }{ \psi }
       +  8 \left( \rho \sigmabar \right)_{,z   }
              \frac{ \psi_{,z     } }{ \psi }
\nonumber\\
   +   {\psi^4 \over 2 \alpha^2}
       \left[  (  \beta^\rho{}_{,\rho} - \beta^z{}_{,z}      )^2
             + (  \beta^z{}_{,\rho}    + \beta^{\rho}{}_{,z} )^2 \right]
   + {\psi^4 \over 6\alpha^2}
        \left[ 2\alpha\rho\bar{\Omega}
          + \beta^\rho{}_{,\rho}
          - \beta^z{}_{,z}   \right]^2
\nonumber\\
 = 
            - 16 \pi \left( \Pi^2 + \Phi_{,\rho}{}^2 + \Phi_{,z}{}^2 \right)
            -  6 \left( \rho^2 \left(\rho \sigmabar\right)_{,\rho} \right)_{,\rho^3}
\nonumber\\
            -  2 \left( \left( \rho \sigmabar \right)_{,\rho} \right)^2
            -  2 \left( \rho \sigmabar \right)_{,zz}
            -  2 \left( \left( \rho \sigmabar \right)_{,z   } \right)^2 
\label{hc_eqn}
\end{eqnarray}
The $\rho$ and $z$ momentum constraints are
\begin{eqnarray}
   \frac{2}{3}\beta^{\rho}{}_{,\rho\rho}
 + \beta^{\rho}{}_{,zz}
 + \frac{1}{3} \beta^z{}_{,z\rho}
 + 32\pi {\alpha \over \psi^2} \Pi_{,\rho}
 - \left(    \left(    \frac{ \alpha_{,z} }{ \alpha }
                   - 6 \frac{ \psi_{,z}   }{ \psi   } \right)
           - \left( \rho\sigmabar                   \right)_{,z}
         \right)
   \left(    \beta^z{}_{,\rho}
           + \beta^\rho{}_{,z}
         \right)
\nonumber\\
 - \frac{2}{3} \left(     \frac{ \alpha_{,\rho} }{ \alpha }
                      - 6 \frac{ \psi_{,\rho}   }{ \psi   }
                     \right)
     \left[
               \beta^\rho{}_{,\rho}
             - \beta^z{}_{,z}
    \right]
 - \frac{8}{3} \alpha \bar{\Omega}
 -\frac{2 \alpha \rho }{ 3 } \left[
                                     6 \omegabar \frac{\psi_{,\rho}}{ \psi }
                                   + \omegabar_{,\rho}
                                   + 3 \omegabar \left(\rho \sigmabar   \right)_{,\rho}
                                   \right]=0,
\label{brho_eqn}
\end{eqnarray}
and
\begin{eqnarray}
   \beta^z{}_{,\rho\rho}
 + \frac{4}{3}\beta^z{}_{,zz}
 - \frac{1}{3} \beta^{\rho}{}_{,z\rho}
   + \left[
        \left( \rho \sigmabar \right)_{,\rho}
      + \frac{ 2 \alpha }{ \psi^6} \left(
                                           \frac{\rho \psi^6}{\alpha}
                                         \right)_{,\rho^2}
     \right]
       \left( \beta^\rho{}_{,z} + \beta^z{}_{,\rho} \right)
\nonumber\\
   + \left[   2 \left( \rho \sigmabar \right)_{,z}
            - \frac{4}{3} \left(    \frac{ \alpha_{,z} }{ \alpha }
                                - 6 \frac{ \psi_{,z}   }{ \psi   } \right)
           \right]
       \left( \beta^z{}_{,z} - \beta^\rho{}_{,\rho} \right)
 - \frac{2 \alpha\rho}{3} \left(
                                  6 \omegabar \frac{\psi_{,z}}{\psi}
                                + \omegabar_{,z}
                               \right)
\nonumber\\
 + 32\pi {\alpha \over \psi^2} \Pi_{,z}
 - 2\alpha(\sigmabar_{,z})\rho^2\omegabar=0
\label{bz_eqn}
\end{eqnarray}
The definition of $\omegabar$ (\ref{omegabar_def}) gives an 
evolution equation for $\sigmabar$:
\begin{equation}\label{sigmabar_eqn}
\sigmabar_{,t} =
        2 \beta^{\rho} \left( \rho\sigmabar \right)_{,\rho^2} + \beta^z
        \sigmabar_{,z}
     -  \alpha\bar{\Omega}
     - \left[ {\beta^{\rho} \over \rho} \right]_{,\rho}
\end{equation}
The evolution equation for $\omegabar$ is
\begin{eqnarray}
\bar{\Omega}_{,t} =
     2 \beta^{\rho} \left(\rho\bar{\Omega}\right)_{,\rho^2}
    +  \beta^z \bar{\Omega}_{,z}
    - {1\over 2\alpha\rho} \left(
                                   \beta^{z}{}_{,\rho}{}^2
                                 - \beta^{\rho}{}_{,z}{}^2
                                 \right)
    + {1\over \psi^4} \left( {\alpha_{,\rho} \over \rho} \right)_{,\rho}
\nonumber\\
    + {\alpha \over \psi^6}\left(
               {(\psi^2)_{,\rho} \over \rho} \right)_{,\rho}
    - {2\alpha \over \psi^4}
        \left(   4 \frac{ \psi_{,\rho^2} }{ \psi }
               + \left( \rho \sigmabar \right)_{,\rho^2}
              \right)
        \left(   \frac{\alpha_{,\rho}    }{ \alpha }
               + \frac{ 2\psi_{,\rho} }{ \psi   }
              \right)
\nonumber\\
- {\alpha \over \psi^4} \left[
                                 \sigmabar_{,z} \left(
                                           \frac{   \alpha_{,z} }{ \alpha }
                                         + \frac{  2\psi_{,z}   }{ \psi   }
                                         \right)
                               + \rho \sigmabar_{,z}{}^2
                               + \sigmabar_{,zz}
                               \right]
    + 64\pi{\alpha\over \psi^4} \rho (\Phi_{,\rho^2})^2
\label{omegabar_eqn}
\end{eqnarray}
The definition of $\Pi$ and the wave equation for $\Phi$ give
\begin{equation}\label{phi_eqn}
\Phi_{,t} =
       \beta^{\rho} \Phi_{,\rho} + \beta^z \Phi_{,z}
            + {\alpha \over \psi^2 } \Pi,
\end{equation}
and
\begin{eqnarray}
{\Pi}_t =
       \beta^{\rho} \Pi_{,\rho} + \beta^z \Pi_{,z}
     + {1\over3} \Pi \left(   \alpha\rho\bar{\Omega}
                            + 2 \beta^{\rho}_{,\rho}
                            + \beta^z_{,z}
                     \right)
\nonumber\\
     + \frac{1}{\psi^4} \left[
                         2\left( \rho \alpha \psi^2 \Phi_{,\rho} \right)_{,\rho^2}
                        + \left(      \alpha \psi^2 \Phi_{,z}    \right)_{,z}
                              \right]
     + \frac{\alpha}{\psi^2} \left[
                                     \left( \rho \sigmabar \right)_{,\rho} \Phi_{,\rho}
                                   + \left( \rho \sigmabar \right)_{,z   } \Phi_{,z}
                                  \right]
\label{pi_phi_eqn}
\end{eqnarray}

The maximal slicing condition $K=0$ gives an independent 
hyperbolic evolution equation for $\psi$:
\begin{equation}\label{psi_evo_eqn}
\dot \psi =  \psi_{,z}\beta^z+\psi_{,\rho}\beta^\rho+
             \frac{\psi}{6}\left(2\beta^\rho_{,\rho}+\beta^z_{,z}+
             \rho\alpha\bar{\Omega}\right).
\end{equation}
\subsection{Boundary Conditions}
The set of axis regularity conditions, applied at $\rho=0$ are:
\begin{eqnarray}
\alpha_{,\rho} & = & 0, \nonumber\\
\psi_{,\rho} & = & 0, \nonumber\\
\bz_{,\rho} & = & 0, \nonumber\\
\brho & = & 0, \nonumber\\
\sigmabar & = & 0, \nonumber\\
\omegabar & = & 0, \nonumber\\
\Phi_{,\rho} & = & 0, \ \ \ \nonumber\\
\Pi_{,\rho} & = & 0. \label{ibc_sum}
\end{eqnarray}
The outer boundary conditions used, applied at $\rho=\rho_{max}$,
$z=z_{max}$ and $z=z_{min}$, are:
\begin{eqnarray}
\alpha &=& 1, \nonumber\\
\psi-1 + \rho \psi_{,\rho} + z \psi_{,z} &=& 0 , \nonumber\\
\bz &=& 0,
\nonumber\\
\brho &=& 0 , \nonumber\\
r \sigmabar_{,t} + \rho \sigmabar_{,\rho} + z \sigmabar_{,z} + \sigmabar&=&0
, \nonumber\\
r \omegabar_{,t} + \rho \omegabar_{,\rho} + z \omegabar_{,z} + \omegabar&=&0
, \nonumber\\
r \Phi_{,t} + \rho \Phi_{,\rho} + z \Phi_{,z} + \Phi&=&0
, \nonumber\\
r \Pi_{,t} + \rho \Pi_{,\rho} + z \Pi_{,z} + \Pi&=&0.
\label{obc_sum}
\end{eqnarray}

\subsection{Finite Difference Operators}\label{app_fde}
In this section, we write out all of the difference operators 
used to convert the differential equations in the previous section
to finite difference equations. At all interior points of the mesh,
the centered forms of the derivate operators are used, and along boundaries,
backward and forward operators are
used as appropriate. Kreiss-Oliger style dissipation 
is applied to evolution equations, at interior points at least two grid points 
inward, in the direction of the stencil, from any boundary.
For $\sigmabar$ and $\omegabar$, we linearly interpolate in $\rho$
at location $\Delta \rho$ (and optionally at $2\Delta \rho$ as well), 
using the values of these variables at $\rho=0$ and $\rho=2\Delta \rho$
(or $\rho=3\Delta \rho$).
Below, we use the notation $u_{i,j}$ to label a point in the mesh corresponding
to coordinate location $((i-1)\Delta \rho,z_{\rm min} + (j-1)\Delta z$
(except for the coordinate variable $\rho$, where it is sufficient to use 
$\rho_i$).
For time derivatives, we use $u^n_{i,j}$ to denote the retarded time level,
and $u^{n+1}_{i,j}$ the advanced time level.
All of the finite-difference operators are $2^{nd}$ order accurate.
\subsubsection{Centered Difference Operators}
\begin{eqnarray}
u_{,\rho} &\rightarrow& 
  \frac{u_{i+1,j}-u_{i-1,j}}{2\Delta\rho} \\
u_{,z}    &\rightarrow& 
  \frac{u_{i,j+1}-u_{i,j-1}}{2\Delta z} \\
u_{,\rho\rho} &\rightarrow& 
  \frac{u_{i+1,j}-2u_{i,j}+u_{i-1,j}}{(\Delta\rho)^2} \\
u_{,zz} &\rightarrow& 
  \frac{u_{i,j+1}-2u_{i,j}+u_{i,j-1}}{(\Delta z)^2} \\
u_{,\rho^2} &\rightarrow& 
  \frac{u_{i+1,j}-u_{i-1,j}}{\rho_{i+1}^2-\rho_{i-1}^2} \\
(u/\rho)_{,\rho} &\rightarrow& 
  \frac{u_{i+1,j}+u_{i,j}}{2\Delta\rho(\rho_{i}+\Delta\rho/2)}-
  \frac{u_{i,j}+u_{i-1,j}}{2\Delta\rho(\rho_{i}-\Delta\rho/2)} \\
\left(u_{,\rho}/\rho\right)_{,\rho} &\rightarrow& 
  \frac{u_{i+1,j}-u_{i,j}}{(\Delta\rho)^2(\rho_{i}+\Delta\rho/2)}-
  \frac{u_{i,j}-u_{i-1,j}}{(\Delta\rho)^2(\rho_{i}-\Delta\rho/2)} \\
u_{,t} &\rightarrow& 
  \frac{u^{n+1}_{i,j}-u^n_{i,j}}{\Delta t} 
\end{eqnarray}

\subsubsection{Forward-Difference Operators}
\begin{eqnarray}
u_{,\rho} &\rightarrow&
   \frac{-u_{i+2,j}+4u_{i+1,j}-3u_{i,j}}{2\Delta\rho} \\
u_{,z} &\rightarrow&
   \frac{-u_{i,j+2}+4u_{i,j+1}-3u_{i,j}}{2\Delta z} 
\end{eqnarray}

\subsubsection{Backward-Difference Operators}
\begin{eqnarray}
u_{,\rho} &\rightarrow&
   \frac{u_{i-2,j}-4u_{i-1,j}+3u_{i,j}}{2\Delta\rho} \\
u_{,z} &\rightarrow&
   \frac{u_{i,j-2}-4u_{i,j-1}+3u_{i,j}}{2\Delta z} 
\end{eqnarray}

\subsubsection{Dissipation Operators}
The following dissipation operator is applied in the $\rho$ direction:
\begin{equation}
\frac{\epsilon_d}{16}\left(u_{i-2,j}-4u_{i-1,j}+6u_{i,j}-
                           4u_{i+1,j}+u_{i+2,j} \right)
\end{equation}
and in the $z$ direction:
\begin{equation}
\frac{\epsilon_d}{16}\left(u_{i,j-2}-4u_{i,j-1}+6u_{i,j}-
                           4u_{i,j+1}+u_{i,j+2} \right),
\end{equation}
where $0 \le \epsilon_d \le 1$.  

\section{Additional Algorithm Details}\label{algorithm_details}
This appendix contains descriptions of a few additional features of the
AMR algorithm described in the paper: computing TE estimates
using a self-shadow hierarchy, details of how the multigrid algorithm
is applied to an adaptive hierarchy, and a description of the 
particular set of interpolation and restriction operators used. 

\subsection{A Self-shadow Hierarchy for Computing Truncation Error Estimates}\label{sec_ss}

A self-shadow hierarchy is a simplification of the idea of using a 
{\em shadow hierarchy} to do truncation error estimation.
A shadow hierarchy is a coarsened (usually with $\rhosp=2:1$ version
of the main hierarchy. Both hierarchies are evolved simultaneously, and
the function values of a given grid in the shadow hierarchy are replaced
with those of the corresponding grid in the main hierarchy whenever the
two are in sync. For example, with $\rhot=2:1$, 
each time step of a shadow grid corresponds to two time steps of the
main grid, and the shadow is updated every two main-grid time steps.
A TE estimate can therefore readily be computed 
by comparing function values in the shadow with corresponding values in
the main hierarchy just before the update step. 

Notice however, that within the recursive time-stepping flow of
the Berger and Oliger algorithm, information for computing a TE estimate is 
``naturally'' available prior to the fine-to-coarse grid injection step 
(see Fig. \ref{BO_ts_fig2}). The coarse level $\ell$ is evolved independently of
the fine level $\ell+1$ from $t_0$ to $t_0+\Delta t_c$, where 
$\Delta t_c$ is the coarse level time step. Also, at $t=t_0$ the 
level $\ell$ grid functions are restricted copies of level $\ell+1$ 
grid functions in the region of overlap $O_\ell^{\ell+1}$. Therefore, 
prior to injection at time $t_0+\Delta t_c$, the difference in an evolved
variable $f$ in levels $\ell$ and $\ell+1$, within the region 
$O_\ell^{\ell+1}$, can serve as an approximation to the truncation error 
$\tau_s(f_{\ell+1})$ for $f$ at level $\ell+1$:\footnote{In fact, if there are only 
two levels in the hierarchy, then this estimate is exactly the truncation
error estimate one would obtain with a shadow hierarchy. 
If levels finer than $\ell+1$ exist,
then in the overlap $O_{\ell+1}^{\ell+2}$ the estimate (\ref{tau_s_def}) will
be modified by an amount of order $(\Delta t_c/\rhosp)^2$.}
\begin{equation}\label{tau_s_def}
\tau_s(f_{\ell+1}) \equiv f_{\ell+1} - f_\ell.
\end{equation}
Therefore, for levels $\ell>1$, one can use (\ref{tau_s_def}) as the
basis for computing truncation error estimates, without the need to refer
to a shadow hierarchy (i.e., the main hierarchy ``casts its own
shadow'', hence the name self-shadow hierarchy). This
method cannot give a TE estimate for the coarsest level (1) in the hierarchy,
and so we require that the coarsest level always be fully refined. 
Thus, the resolution of level 2 is chosen to match the desired
coarsest resolution for a given problem (for the sample evolutions described in
in Sec. \ref{sec_results}, level 2 is always quoted as the base level).

In practice, a slightly modified form of $\ref{tau_s_def}$ is used,
as we now describe. Depending upon the problem, one or more 
of $\psi$, $\Phi$, $\Pi_\Phi$, $\omegabar$, and
$\sigmabar$ are used in the calculation (i.e., all evolved quantities---$\psi$ is
not used when the Hamiltonian constraint is used to solved for $\psi$ in 
a fully constrained evolution).
Optionally, the truncation error estimate is scaled by the norm of the function, if 
the norm is larger than some constant $k$ ($k=1$ typically), and can also be 
multiplied $\rho^p$, for some integer $p$ 
chosen heuristically to either enhance or reduce the near-axis refinement.
The TE estimate for a function $f_\ell$ at level $\ell$ is thus
defined to be
\begin{equation}\label{ref_tre}
\tau_s(f_\ell)=\frac{\left(f_\ell-f_{\ell-1}\right) \rho^p}
                    {{\rm max}(k,||f_\ell||_2)},
\end{equation}
where it is implied that $f_\ell$ is only defined in the overlap between levels
$\ell$ and $\ell-1$, $f_\ell$ is restricted to the resolution of level
$\ell-1$ prior to subtraction, and the result is then interpolated back to
the resolution of level $\ell$. Typically, we use $p=2$ for $\omegabar$,
$p=1$ for $\sigmabar$, and $p=0$ for the other variables. 
The TE estimate for the level is defined to be
\begin{equation}\label{ref_tre_l}
\tau_s(\ell)=\sqrt{\sum{\tau_s(f_\ell)^2}},
\end{equation}
where the sum is taken over the desired subset of variables listed
above. 

Optionally, the TE estimate calculated in (\ref{ref_tre_l}) is further smoothed 
(using simple averaging over a 5-by-5 square cell of points), and/or 
scaled by a quantity $H_\ell\geq 1$ in the region of overlap between levels 
$\ell$ and $\ell+1$. $H_\ell$ therefore provides a kind of 
``hysteresis'' to the truncation error estimation process: when the TE estimate in a 
region of level $\ell$ grows above $\tau_{\rm max}$, that region is refined;
however, for the region to be unrefined at a later time, the TE estimate needs to drop 
below $\tau_{\rm max}/H_\ell$ there.
In most of the simulations we keep $H_\ell=1$, though occasionally it has proven 
useful to set it to around $5-10$.

\subsection{Multigrid on an Adaptive Hierarchy}\label{sec_mg_ad}

The FAS multigrid algorithm, with $V$-cycling, that we use to solve elliptic
equations on a grid hierarchy such as that shown in Fig. \ref{BO_mesh_struct} is 
based on Brandt's multi level adaptive (MLAT) scheme \cite{brandt}, and is
similar to that used in \cite{brown_lowe2}. 
To simplify the algorithm,
we require that $\rhosp = 2^q$ for some integer $q$; then the AMR hierarchy
can easily be extended to incorporate the multigrid levels, which have a refinement
ratio $\rho_{\rm mg}$ of 2:1. 
When building the multigrid hierarchy, each AMR grid is individually coarsened by
factors $\rho_{\rm mg}$ (i.e. factors of 2) until either a) one dimension of the grid is smaller
that the minimum allowed, or b) the coarsened grid can be ``absorbed'' into
a larger grid at that level in hierarchy. With this type of hierarchy 
the $V$-cycle
begins with relaxation of the finest level grids only. Then as the finer levels are 
coarsened they are absorbed, if possible, into coarser levels. This method of
relaxation on an adaptive hierarchy is in contrast to the Fast Adaptive Composite
Grid Method\cite{mccormick}, though we do not believe that 
the particular details of how the elliptic equations are solved have 
much bearing on the algorithm described here for
dealing with coupled elliptic/hyperbolic equations.

A couple of minor points regarding the details of the multigrid solution are 
worth mentioning.
First, inter-level operations, such as 
restriction, computing coarse-grid corrections, etc., are only performed in 
the region of overlap between the two levels (which is {\em always} the region
of the fine level, given the kind of hierarchies that are produced by B\&O AMR).
Second, the manner in which the relaxation sweep proceeds over a level is
modified, to account for possible grid overlap\footnote{We require
that grids at the same level {\em must} overlap when spanning a connected region
of the computational domain. In other words, it is not sufficient for grids
to merely ``touch'' at a common boundary between them. This requirement simplifies the
relaxation subroutine so that it can operate locally on a grid-by-grid
basis, without needing to communicate adjacent information. However, as described
in the text, the communication step then needs to be shifted to other
parts of the algorithm.}, as follows.
During the relaxation sweep, the variables at a given coordinate location and
level are relaxed {\em only once}, regardless of the number of grids encompassing that
location. This is crucial in order to 
preserve the smoothing properties of the relaxation scheme. We use a mask function 
to enforce this requirement of a single update per grid point. The mask is initialized to zero 
on all grids at that level, prior to a relaxation sweep. Then, on a given level, a sweep is
applied, in turn, to each grid in the level, but only variables at points
where the mask is equal to zero are modified.
After the sweep is complete on a given grid, the mask is set to one throughout the
grid, and the mask and other grid functions are 
copied to overlapping grids at the same level. Therefore, subsequent relaxation 
sweeps on adjacent grids skip over points that have already been relaxed.
This communication step, in addition to enforcing a single update per point,
ensures that grid functions
are numerically unique at all physical grid locations\footnote{Note that 
such a communication step is also performed after relaxation of evolved
variables, during the Crank-Nicholson iteration.}, which 
is important for preserving the convergence properties of multigrid.
Also, although Dirichlet boundary conditions are used
at interior boundaries (i.e. those not abutting boundaries of the computational domain) of any single grid, 
the communication ensures 
that points interior to a union of grids are ultimately updated using the PDEs, 
even if they lie on the boundary of some grid in the overlap region.

\begin{figure}
\begin{center}
\includegraphics[width=14cm,clip=true]{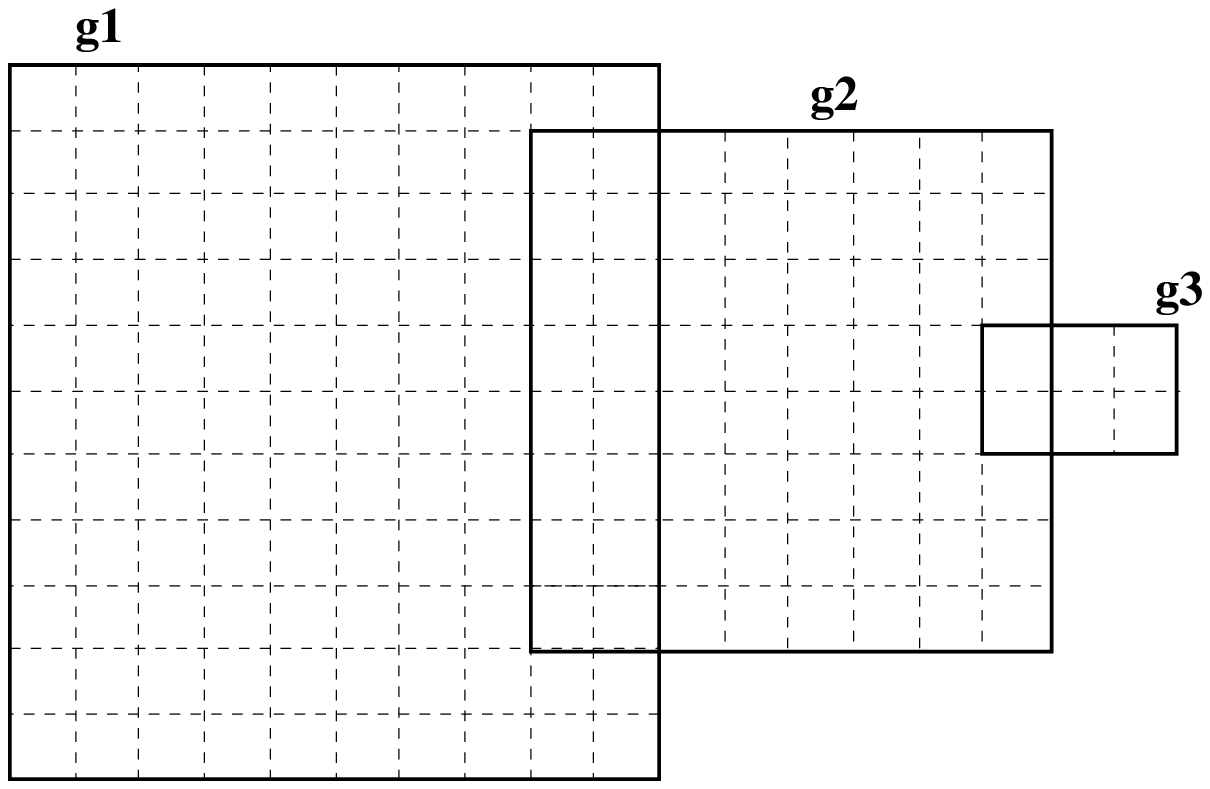}
\end{center}
\caption
[A hypothetical multigrid grid configuration that will adversely affect
execution speed]
{A hypothetical multigrid grid configuration that will adversely affect
execution speed. 
In the figure we depict three grids, {\bf g1,g2} and
{\bf g3}, several levels down in the grid hierarchy (i.e., after 
several coarsening steps have already been performed). Grid {\bf g3}
cannot be coarsened any further, while grids {\bf g1} and {\bf g2}
can, and ideally should be coarsened further, to maintain the speed of
the algorithm. However, because {\bf g3} overlaps the other two
grids, this entire level must be considered a coarsest level,
and solved ``exactly''.}
\label{mg_cprob}
\end{figure}

With regards to the performance of this multigrid scheme on a general adaptive hierarchy,
there are two situations of relevance where performance could suffer, 
compared to
the single grid multigrid algorithm. The first occurs when, at some level down
(coarser) in the multigrid hierarchy, one or more grids in a connected union of grids 
is a ``coarsest grid'', and hence needs to be 
solved ``exactly''\footnote{Here, ``exactly'' means that the residual on the
coarsest grid is reduced by several orders of magnitude by relaxation (it is 
usually not necessary to solve the coarse-grid problem to within machine precision).}--- see Fig. 
\ref{mg_cprob}. Experimentation showed that the {\em entire union}
needs to be solved exactly in that situation; i.e. it is not sufficient to 
solve the equations exactly on the coarsest grids, then proceed down the $V$-cycle 
on the remaining grids.
If the union of grids consists of a relatively small number of grid points, 
then such a situation will not be a problem; otherwise, there will be a 
significant slow-down of the code, for the speed of an exact solve
suffers dramatically as the number of unknowns increase.
To date, we have been able to avoid this potential speed bottleneck
by using a more simplistic clustering method that does not produce
grid-overlap, as discussed in the following section.

The second situation where performance suffers is when the TE estimate requires long,
skinny rectangular regions to be refined. This does occur with the more 
prolate initial data configurations that have been studied \cite{fpthesis}.
What happens then, is that such an elongated grid
can not be coarsened very much along the larger grid dimension before the smaller 
dimension has reached the smallest allowed size. Again, this results 
in a relatively large number of points on the coarsest grid where the solution needs to
be obtained exactly. As of yet this problem has not been addressed.

\subsection{Clustering}\label{sec_cluster}
We have incorporated two clustering routines into the code. The first,
written by Reid Guenther, Mijan Huq and Dale Choi \cite{ref_clusterer},
is based upon the signature-line method of Berger and Rigoutsos 
\cite{BR}. The second is a simple routine that produces
single, isolated clusters---each isolated region of
high TE is surrounded by a single cluster, and then all clusters
within a certain distance of each other are merged together into an
encompassing cluster. For the problems 
studied so far, the isolated cluster method turns out to be almost as
efficient as the signature-line method. Therefore, since efficiency is not an 
issue, the isolated cluster method is preferable, because it avoids one of the 
potential speed-bottlenecks of the multigrid algorithm discussed in 
Section \ref{sec_mg_ad}; furthermore, as mentioned in Section 
\ref{sec_hf_gb_noise}, minimizing cluster overlap helps reduce high-frequency 
noise problems.

A clustering issue that needs to be dealt with in this code is that the
resultant grid hierarchy must be compatible with the multigrid solver.
This places two restrictions on cluster sizes and positions. First,
an individual grid must have dimensions that can be factored into
$x_{\rm min}2^n$, where $x_{\rm min}$ is one of the smallest, allowed grid dimensions, 
and $n$ is a non-negative integer. Second, if several grids overlap,
then their relative positions must be such that the common grid points
align on all possible levels of a multigrid hierarchy. Specifically, if a union of 
overlapping grids can collectively be coarsened $m$ times in the multigrid 
hierarchy, then the relative offsets of grid origins on the finest level must be
multiples of $2^m$ grid points. These requirements are enforced
after the initial clustering algorithm is called, by modifying the returned
cluster list accordingly. This gives more flexibility to experiment
with different clustering routines, which consequently do not need to be aware
of the alignment issues.

To conclude this section we mentioned a couple of additional options that have
been implemented in the post-clustering routine. 
They are adding ``ghost zones'' between
adjacent, touching clusters, so that both the multigrid and evolution relaxation sweeps
correctly solve the system of equations in a domain given by the 
union of grids at a given level; and optionally moving or shrinking clusters,
if necessary, to prevent them from touching parent boundaries\footnote{In 
principle this should never occur if one adds a buffer zone about 
the region of high truncation error. However, because of the grid shuffling
performed to obtain a hierarchy acceptable to multigrid, a grid could be extended to
touch a parent boundary. With the option enabled to prevent this, the
grid will be reduced rather than extended to fit into the multigrid scheme. This
comes at the expense of not obtaining ``optimal'' zones about the region of
high truncation error.},
which helps to avoid instabilities that occasionally occur in such
situations.

\subsection{Interpolation and Restriction Operators}\label{sec_ip_ops}
Here we state the restriction and interpolation operators used in the AMR
code. Simple injection is used to restrict a fine grid to a coarse grid
during the injection phase of the AMR algorithm, and when computing the
TE estimate. A fourth order (bi-cubic) interpolation scheme is used to initialize newly
refined fine grids (or regions thereof) from the encompassing coarser grid.
The scheme proceeds by first interpolating every row of the coarse grid 
to the fine grid (i.e. every $\rhosp$th row of the fine grid is filled in),
then all the remaining points on the fine grid are computed via
interpolation, column-by-column.
The multigrid routine uses half-weight restriction when transferring
from a fine to coarse grid, and linear interpolation for the coarse to 
fine transfer.

\subsection{Initializing the Grid Hierarchy}\label{sec_grid_init}
Fig. \ref{init_fig} below contains a pseudo-code description of
the steps used to initialize the grid hierarchy.

\begin{figure}
\epsfxsize=17cm
\begin{flushleft}
\begin{obeylines}
{\tt
~~~~~~t:=0;
~~~~~~initialize the grid hierarchy with $2$ levels, each covering the entire domain; {\em (so $\ell_f=2$)}
~
~~~~~~repeat
~~~~~~~~~$\ell^p_f:=\ell_f$;
~~~~~~~~~call set\_initial\_data(); {\em (see below)}
~~~~~~~~~call set\_past\_t\_data\_1st\_order(); {\em (see below)}
~~~~~~~~~call single\_step(1); {\em (see Fig.\ref{BO_ts_fig2})}
~~~~~~~~~regrid the entire hierarchy using truncation error estimates
~~~~~~~~~~~~computed in the previous step; {\em (thus possibly changing $\ell_f$)}
~~~~~~~~~reset $t$ to $t:=0$ while retaining the current hierarchy structure;
~~~~~~until $\ell_f=\ell^p_f$ or maximum number of refinement levels reached
~
~~~~~~call set\_initial\_data();
~~~~~~call set\_past\_t\_data\_1st\_order();
~
~~~~~~call single\_step(1); {\em (evolve hierarchy {\bf forwards} in time one coarse step)} 
~~~~~~call flip\_dt; {\em (see below)}
~
~~~~~~call single\_step(1); {\em (evolve hierarchy {\bf backwards} in time one coarse step)} 
~~~~~~call flip\_dt;
~
~~~~~~call set\_initial\_data();
~
~~~~~~{\em (done computing initial data and hierarchy)}
~
~~~~~~subroutine set\_initial\_data()
~~~~~~~~~initialize hyperbolic variables over levels $[1..\ell_f]$ with freely-specifiable data;
~~~~~~~~~solve the elliptic equations over levels $[1..\ell_f]$;
~~~~~~end of subroutine set\_initial\_data
~
~~~~~~subroutine flip\_dt()
~~~~~~~~~for each elliptic variable $f_{1}(t)$: $f_{1}(t+\Delta t_1):=2 f_{1}(t)-f_{1}(t-\Delta t_1)$;
~~~~~~~~~do $\ell:=2$ to $\ell_f$
~~~~~~~~~~~~for each elliptic variable $f_{\ell}(t)$: $f_{\ell}(t+\rhot\Delta t_\ell):=2 f_{\ell}(t)-f_{\ell}(t-\rhot\Delta t_\ell)$;
~~~~~~~~~~~~$\Delta t_\ell:=-\Delta t_\ell$;
~~~~~~~~~end do
~~~~~~end of subroutine flip\_dt
~
~~~~~~subroutine set\_past\_t\_data\_1st\_order()
~~~~~~~~~do $\ell:=1$ to $\ell_f$
~~~~~~~~~~~~for each elliptic variable $f_{\ell}(t)$: $f_{\ell}(t-\rhot\Delta t_\ell):=f_{\ell}(t)$;
~~~~~~~~~end do
~~~~~~end of subroutine set\_past\_t\_data\_1st\_order
}
\end{obeylines}
\end{flushleft}
\caption
{A pseudo-code description of the steps we use to initialize the grid hierarchy. 
The {\em repeat} loop is used to calculate the hierarchy structure at $t=0$.
Then, to initialize past time level data for elliptic variables using
linear extrapolation (which is done in $\mbox{flip\_dt()}$),
the entire hierarchy is evolved forwards, then backwards in time by single
coarse level time steps (alternatively we could evolve backwards,
then forwards in time here---the results would essentially be the same).
The idea behind this last step is that since the evolution
of the hyperbolic variables is driving any change in the elliptic 
variables, we can use the results of a small evolution step
to provide a better estimate of past time level information than first-order
extrapolation of the solution at $t=0$. In principle, this step can
be iterated if need be, though we found that a single step
is sufficient (though {\em not} always necessary depending on the free initial data) 
to obtain close to second order 
convergence of the final solution.}
\label{init_fig}
\end{figure}

\subsection{Controlling High-Frequency Grid-Boundary
Noise}\label{sec_hf_gb_noise}
An issue that needs to be dealt with in a Berger \& Oliger style AMR
scheme is controlling high-frequency solution components (``noise'') 
that may occur at parent-child  
grid boundaries. For a second order accurate finite-difference scheme, 
the second derivatives of grid functions are typically not continuous 
across the boundaries after child to parent injection. This potential source 
of high-frequency noise on the parent level is rather efficiently eliminated 
by the Kreiss-Oliger dissipation filters we incorporate into our finite 
differenced evolution equations.

\begin{figure}
\begin{center}
\psfrag{im}{Interpolation Method}
\psfrag{i1}{(1) linear in time from parent grids}
\psfrag{i2}{(2) $4^{th}$ order in space using points from (1)}
\psfrag{i3}{(3) $4^{th}$ order in space using a point from (1) or (2)}
\psfrag{i3b}{and 3 interior points normal to the boundary}
\psfrag{i4}{(4) as (3), though utilizing points in (3) instead of}
\psfrag{i4b}{interior points}
\psfrag{i5}{(5) bilinear in space using points from (1) and (4)}
\includegraphics[width=14cm,clip=true]{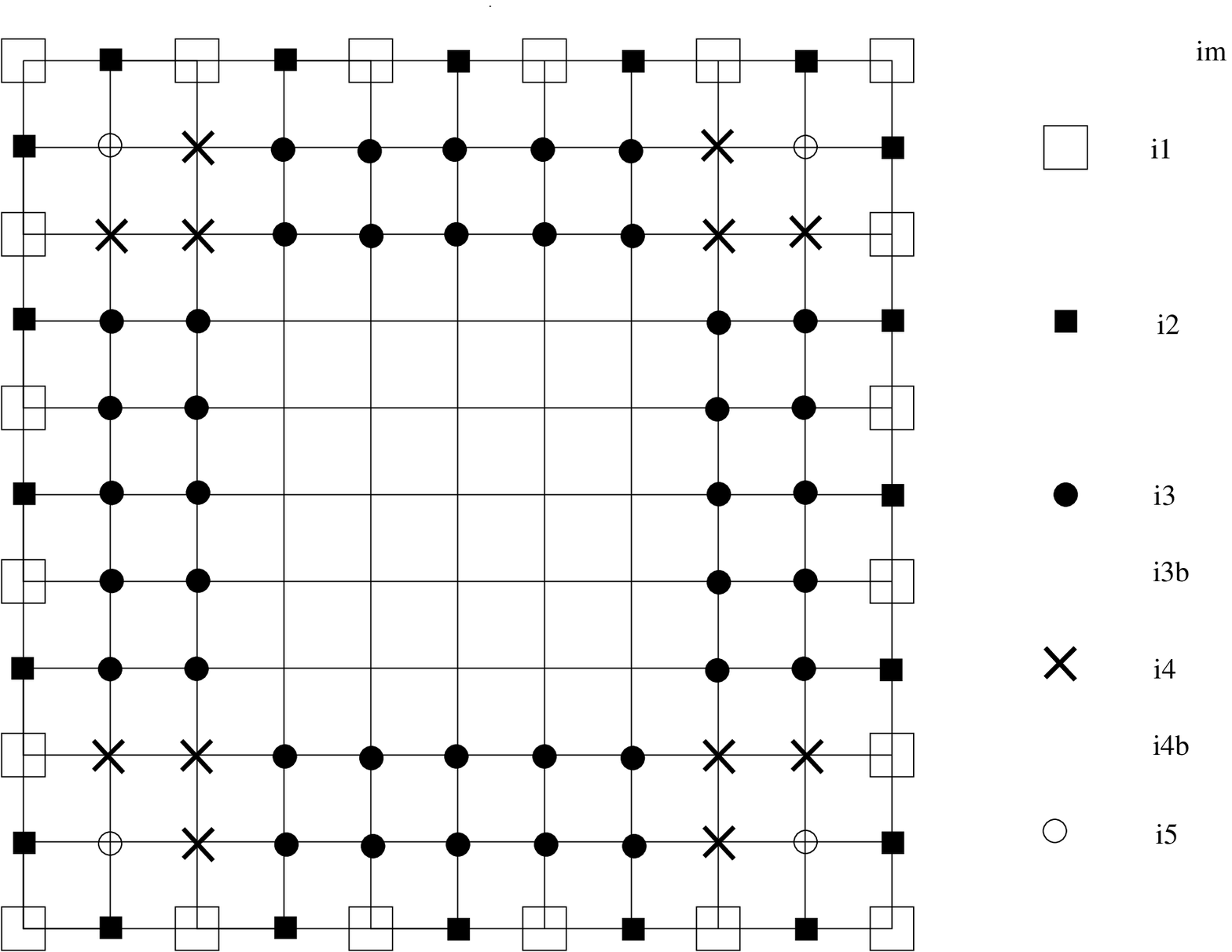}
\end{center}
\caption[An illustration of the interpolation method used for $\sigmabar$ and
$\omegabar$ during AMR evolution]
{An illustration of the interpolation method used for $\sigmabar$ and
$\omegabar$ during AMR evolution.
In the figure we assume that the spatial refinement ratio is $2:1$, and
that all four grid boundaries are AMR boundaries. Points labeled by
(1) and (2) are set once prior to the Crank-Nicholson (CN) iteration, while
points labeled by (3), (4) and (5) are reset after every CN step
(see the pseudo-code in Fig. \ref{pseudo_bar_interp}). Points
not explicitly labeled are ``interior'' points, and are evolved.
\label{fig_bar_interp}}
\end{figure}

In certain situations we have found that high-frequency noise also develops
on child grids, within a grid point or two of the AMR boundary (in particular
near the corners of the grid, or places where two grids overlap). This
noise is not as easily dealt with, as the Kreiss-Oliger filter acting
normal to the boundary is only applied a distance three points and further 
away from the boundary. The source of this noise appears to be the parent-child 
interpolation scheme used to set the boundary values, and in general the 
interpolation method must be tailored to each variable in order to reduce the noise
to an acceptable level.
For out current model, we use the following interpolation method. For all evolved
variables ($\sigmabar, \omegabar, \Phi, \Pi_\Phi$ and $\psi$) we use
linear interpolation in time from the parent level to set boundary values 
on the child grid at points coincident with parent grid points. This is
followed by fourth-order interpolation in space (as described
in Section \ref{sec_ip_ops}) for the remaining boundary points. 
Furthermore (see Fig. \ref{fig_bar_interp} and \ref{pseudo_bar_interp}),
after each step of the Crank-Nicholson iteration we reset 
$\omegabar$ and $\sigmabar$ in a zone two grid points in from AMR boundaries with values
obtained either 1) by fourth order interpolation using function values
from the boundary and three additional points inward from this zone,
or 2) via bilinear interpolation at ``corner'' points, i.e. those points that
are a single cell width away from two boundaries. This 
technique for $\sigmabar$ and $\omegabar$ was discovered after quite
a bit of experimentation with different interpolation schemes, and is 
quite effective in reducing the level of noise at the grid boundaries.

\begin{figure}
\epsfxsize=17cm
{
\begin{verbatim}
subroutine interp_interior_AMR_bnd(grid function f[1..Nrho,1..Nz])
   for i=3 to Nrho-2 do
      f(i,2)=0.4*f(i,1)+2.0*f(i,4)-2.0*f(i,5)+0.6*f(i,6)
      f(i,3)=0.1*f(i,1)+2.0*f(i,4)-1.5*f(i,5)+0.4*f(i,6)
      f(i,Nz-1)=0.4*f(i,Nz)+2.0*f(i,Nz-3)-2.0*f(i,Nz-4)+0.6*f(i,Nz-5)
      f(i,Nz-2)=0.1*f(i,Nz)+2.0*f(i,Nz-3)-1.5*f(i,Nz-4)+0.4*f(i,Nz-5)
   end do

   for j=3 to Nz-2 do
      f(2,j)=0.4*f(1,j)+2.0*f(4,j)-2.0*f(5,j)+0.6*f(6,j)
      f(3,j)=0.1*f(1,j)+2.0*f(4,j)-1.5*f(5,j)+0.4*f(6,j)
      f(Nrho-1,j)=0.4*f(Nrho,j)+2.0*f(Nrho-3,j)-2.0*f(Nrho-4,j)+0.6*f(Nrho-5,j)
      f(Nrho-2,j)=0.1*f(Nrho,j)+2.0*f(Nrho-3,j)-1.5*f(Nrho-4,j)+0.4*f(Nrho-5,j)
   end do

   f(2,2)=(f(1,1)+f(3,3)+f(1,3)+f(3,1))/4
   f(Nrho-1,2)=(f(Nrho,1)+f(Nrho,3)+f(Nrho-2,1)+f(Nrho-2,3))/4
   f(2,Nz-1)=(f(1,Nz)+f(3,Nz)+f(1,Nz-2)+f(3,Nz-2))/4
   f(Nrho-1,Nz-1)=(f(Nrho,Nz)+f(Nrho,Nz-2)+f(Nrho-2,Nz)+f(Nrho-2,Nz-2))/4

end of subroutine interp_interior_AMR_bnd
\end{verbatim}
}
\caption[A pseudo-code description of part of the interpolation method
used for $\sigmabar$ and $\omegabar$ during AMR evolution]
{ A pseudo-code description of part of the interpolation method
used for $\sigmabar$ and $\omegabar$ during AMR evolution. For simplicity,
in this subroutine we assume that all four boundaries are interior to
the computational domain boundaries. The set of points altered here correspond
to (3), (4) and (5) in Fig. \ref{fig_bar_interp}, and the interpolation
operators used are independent of the spatial refinement ratio (as opposed
to points (1) and (2) in Fig. \ref{fig_bar_interp}).
\label{pseudo_bar_interp}}
\end{figure}

For the elliptic variables ($\alpha, \brho$, $\bz$ and optionally $\psi$), prior to a 
Crank-Nicholson evolution cycle, we reset these
variables on AMR boundaries at points unique to the grid 
(in between points coincident with parent level points---those 
labeled by (2) in Fig. \ref{fig_bar_interp}) using fourth
order interpolation from the remaining points on the boundary. This 
overwrites the values set by
coarse-grid corrections (CGCs) during the most recent multigrid solve 
that involved coarser levels. The reason for doing this is as follows.
In multigrid, CGCs typically
introduce high-frequency noise on the finer level, while the
subsequent post-CGC relaxation sweeps smooth out this noise. However,
since no relaxation is applied on AMR boundaries, some form
of explicit smoothing is required --- the above described fourth
order interpolation provides this smoothing mechanism.

Another stage in the algorithm where high-frequency noise can creep
into the solution is during the regridding phase, if the refined region
on a given level expands. Then, within the part of a new 
grid overlapping the old refined region, grid functions will be initialized
by copying data from an old fine grid, while on the remaining part of the new grid the
grid functions will be initialized via interpolation from a parent 
grid. Sometimes, at the interface between the copied/interpolated data,
tiny discontinuities are introduced.
The grid functions are then easily smoothed
by applying a Kreiss-Oliger filter to them (at all time
levels involved) after regridding.

\end{appendix}

\end{document}